\documentclass[a4paper,10pt,notitlepage]{article}
\usepackage{amsfonts,amssymb,amsmath,amsthm,mathtools}
\usepackage{caption}
\usepackage{cite}
\usepackage{caption}
\usepackage{upgreek}
\usepackage[utf8]{inputenc}
\usepackage{wasysym}
\newcommand{\beqa}{\begin{eqnarray}}
\newcommand{\eeqa}{\end{eqnarray}}
\newcommand{\beq}{\begin{equation}}
\newcommand{\eeq}{\end{equation}}
\DeclareMathOperator{\fc}{\mathfrak{c}}
\DeclareMathOperator{\fa}{\mathfrak{a}}

\newcommand{\fQ}{\mathbf{\mathcal{Q}}}

\newcommand{\SU}{\ensuremath{\text{SU}}}
\newcommand{\SL}{\ensuremath{\text{SL}}}
\newcommand{\UU}{\ensuremath{\text{U}}}
\newcommand{\calS}{\mathcal{S}}

\newcommand{\calZ}{\mathcal{Z}}
\newcommand{\calN}{\mathcal{N}}
\newcommand{\calW}{\mathcal{W}}
\newcommand{\calC}{\mathcal{C}}

\newcommand{\ba}{\boldsymbol{\alpha}}
\newcommand{\bb}{\boldsymbol{\beta}}

\newcommand{\car}{{\textbf{car}}}
\newcommand{\shap}{\textsf{Q}}
\newcommand{\ver}{\textsf{V}}
\newcommand{\gv}{\gamma}
\newcommand{\gvJ}{\boldsymbol{\upgamma}}
\newcommand{\gvt}{\bar{\gamma}}
\newcommand{\block}{\mathcal{B}}
\newcommand{\e}{\operatorname{e}}
\newcommand{\pro}{\pi}
\newcommand\id{{\textbf{I}}}%{\mathbbm{1}}
\newcommand{\bracket}[2]{\ensuremath{\left< #1\, |\, #2\right>}}

\newcommand{\vac}[1]{\ensuremath{\left< \, #1\, \right>}}
\newcommand{\vvac}[1]{\ensuremath{\left<\left< \, #1\, \right>\right>}}

\newcommand{\com}[2]{\ensuremath{\left[ #1\, ,\, #2\right]}}

\newcommand{\form}[2]{\ensuremath{\left( #1, #2\right)}}

\newcommand{\bW}{\textbf{W}}
\newcommand{\bw}{\textbf{w}}
\newcommand{\bY}{\textbf{Y}}

\newcommand{\cd}{\Delta} % conformal dimensions
\newcommand{\charge}{w} % W-charges
\newcommand{\bcharge}{\textbf{w}} % W-charges
\newcommand{\cuco}{\phi} % Coefficients of the curve 
\newcommand{\hs}{\textsf{h}} % weights of the SU(N) fundamental representation
\newcommand{\bphi}{\boldsymbol{\varphi}} % vector of free 2D bosons (for the SU(N) Toda representation at c=N-1 through free bosons
\def\balpha{\ba}\def\bbeta{\bb} % for the weights in the Toda theory
\newcommand{\fb}{\lambda} % The free boson before the orbifold
\newcommand{\define}{\stackrel{\text{def}}{=}}
\newcommand{\mL}[1]{\ensuremath{m_{L,\,#1}}}
\newcommand{\mR}[1]{\ensuremath{m_{R,\,#1}}}
\newcommand{\mLt}[1]{\ensuremath{\tilde{m}_{L,\,#1}}}
\newcommand{\mRt}[1]{\ensuremath{\tilde{m}_{R,\,#1}}}
\newcommand{\am}{\textsf{a}} % The modes of the  free boson field

\begin{document}
%----------------------------------------------------------------------------------------------

%----------------------------------------------------------------------------------------------
%TITLE PAGE 
%----------------------------------------------------------------------------------------------
\thispagestyle{empty}
\setcounter{page}{0}
\begin{flushright}\footnotesize
\texttt{DESY 17-029}\\
\texttt{MITP/17-012}\\
\vspace{0.5cm}
\end{flushright}
\setcounter{footnote}{0}

\begin{center}
{\huge{
\textbf{2D CFT blocks for the 4D class $\mathcal{S}_k$ theories
}
}}
\vspace{15mm}

{\sc Vladimir Mitev$^{a}$,   Elli Pomoni$^{b}$ }\\[5mm]

{\it $^a$Institut f\"ur Physik, WA THEP\\
Johannes Gutenberg-Universit\"at Mainz\\
Staudingerweg 7, 55128 Mainz, Germany
}\\[3mm]

{\it $^b$DESY Hamburg, Theory Group, \\
Notkestrasse 85, D--22607 Hamburg, Germany
}\\[3mm]

\texttt{vmitev@uni-mainz.de}\\
\texttt{elli.pomoni@desy.de}\\[10mm]

\textbf{Abstract}\\[2mm]
\end{center}

 This is the first in a series of papers on the search for the 2D CFT description of a large class of 4D $\mathcal{N}=1$ gauge  theories. Here, we identify the 2D CFT symmetry algebra and its representations, namely the conformal blocks of the Virasoro/W-algebra, that underlie the 2D theory and reproduce the Seiberg-Witten curves of the $\calN=1$ gauge  theories. We find that the blocks corresponding to the $\SU(N)$ $\mathcal{S}_k$ gauge theories involve fields in certain non-unitary representations of the $\bW_{kN}$ algebra. These conformal blocks give a prediction for the instanton partition functions of the 4D $\mathcal{N}=1$ SCFTs of class $\mathcal{S}_k$.

 \newpage
   
 %----------------------------------------------------------------------------------------------
%END OF TITLE PAGE 
%----------------------------------------------------------------------------------------------

\setcounter{page}{1}

%----------------------------------------------------------------------------------------------

\tableofcontents
\addtolength{\baselineskip}{5pt}

%----------------------------------------------------------------------------------------------
%Main text
%----------------------------------------------------------------------------------------------

%----------------------------------------------------------------------------------------------
\section{Introduction}
%----------------------------------------------------------------------------------------------

The study of supersymmetric gauge theories was revolutionized by Seiberg and collaborators in the nineties through the use of holomorphicity, symmetries as well as asymptotics (weak coupling behavior) \cite{Intriligator:1995au}.
Building up on these developments, Seiberg and Witten realized \cite{Seiberg:1994rs,Seiberg:1994aj} that by adding electromagnetic duality (S-duality) to the game, one can obtain the low energy BPS spectrum of $\calN=2$ gauge theories by deriving a holomorphic algebraic curve, the so-called Seiberg-Witten (SW)  curve,  that incorporates all the symmetries (including S-duality) and weak coupling behavior. Soon after, Intriligator and Seiberg \cite{Intriligator:1994sm} obtained the first examples of algebraic curves that compute the low energy coupling constants in the abelian Coulomb phase for $\calN=1$ theories.

In the last decade, the most modern developments in the field  are based on the deep connection of S-duality in 4D gauge theory with 2D modular invariance. In the prototypical example of the maximally supersymmetric  $\calN=4$ super Yang-Mills (SYM), the Montonen-Olive $\SL(2,\mathbb{Z} )$ duality  can be geometrically realized as the modular group of a torus by compactifying the 6D $(2, 0)$ SCFT on a torus  \cite{Vafa:1997mh}.  Similarly, a large class of  4D $\calN=2$ superconformal field theories (SCFTs)s, referred to as  class $\calS$ \cite{Gaiotto:2009we,Gaiotto:2009hg}, can be obtained via compactification of (a twisted version of) the 
6D $(2, 0)$ SCFT on Riemann surfaces of genus $g$ and with $n$ punctures. The parameter space of the exactly marginal gauge couplings is identified with the complex structure moduli space  of the Riemann surface.  What is more,  the partition function of the 4D $\calN=2$ theories on a four sphere\footnote{Technically \cite{Hama:2012bg}, on an ellipsoid with deformation parameter $b^2=\tfrac{\epsilon_1}{\epsilon_2}$, where the $\epsilon_i$ are the $\Omega$-background deformation parameters entering the Nekrasov partition functions. } \cite{Pestun:2007rz} are equal to correlation functions of the 2D Liouville/Toda CFT on  that Riemann surface  \cite{Alday:2009aq,Wyllard:2009hg}, which is the core of the celebrated AGT(W) correspondence. 
The 4D/2D interplay was originally  discovered  for the  $\calN=2$  class $\calS$ theories in \cite{Gaiotto:2009we} by studying the SW curves
and realizing that they arise from the compactification of M5-branes on Riemann surfaces decorated with punctures.  See \cite{Teschner:2014oja,Pestun:2016jze} for recent reviews.

 Motivated by the above developments for  $\calN=2$ theories, we wish to explore how much mileage we can get for theories with only $\calN=1$ supersymmetry. We begin by recalling that it is not uncommon  to find exactly marginal couplings also in  $\calN=1$ supersymmetric theories \cite{Leigh:1995ep,Green:2010da}, with the AdS/CFT correspondence offering a natural route to several examples of $\calN=1$ orbifold daughters of $\calN=4$ SYM \cite{Kachru:1998ys,Lawrence:1998ja}. A very large class of 4D  $\calN=1$ SCFTs, naturally called $\calS_{\Gamma}$ \cite{Heckman:2016xdl,Apruzzi:2016nfr}, arise from M5-branes probing the $\mathbb{C}^2/\Gamma$ ADE singularity. Their study was originated in \cite{Gaiotto:2015usa}, with the $\calS_{k}$ class arising after compactification of $\mathbb{Z}_k$ orbifolds of the (2,0) theory, see also \cite{Franco:2015jna,Hanany:2015pfa} and \cite{Morrison:2016nrt, Heckman:2016xdl, Razamat:2016dpl, Pal:2016eqy, Garcia-Etxebarria:2016bpb, Bah:2017gph}.  The  SW curves for the class $\calS_{k}$ theories were derived and studied in  \cite{Coman:2015bqq}, using Witten's M-theory approach \cite{Witten:1997sc}. 
 
 For $\mathcal{N}=2$ theories, the SW curves completely solve the IR theory.
The $\mathcal{N}=2$ supersymmetry and more specifically the $\SU(2)_R$ relates the holomorphic superpotential to the non-holomorphic (in $\mathcal{N}=1$ superspace) K\"ahler part and thus we can obtain the full prepotential. For theories with only $\mathcal{N}=1$ supersymmetry, we can only hope to fix the holomorphic superpotential part. However, there are $\calN=1$ examples for which also the K\"ahler part can be fixed, see for example \cite{Aganagic:2006ex,Aganagic:2008qa}. From a field theory point of view this should be a consequence of an extra global symmetry. For the theories in class $\calS_\Gamma$, we expect more, than for generic $\mathcal{N}=1$theories, due to their rich global symmetries inherited from the orbifold construction.\footnote{ As explained in \cite{Gaiotto:2015usa,Coman:2015bqq}, the $\SU(2)_R$ is broken by the orbifold, but a diagonal $\UU(1)_R$ remains. Moreover, instead of the $\UU(1)_r$ of $\mathcal{N}=2$, a global symmetry $\UU(1) \times \UU(1)^{k-1} \times \UU(1)^{k-1}$ which is heavily constraining the theory.}

\smallskip

 The purpose of this article is to begin the search for the 2D conformal field theories (CFT), whose correlation functions reproduce the partition functions of the  4D  $\calN=1$ SCFTs of class $\calS_{k}$ and in general of class  $\calS_{\Gamma}$. In principle, there is no reason to expect that such a 4D/2D relation exists for $\calN=1$ theories. We adopt here a conservative approach - if such a relation exists, then the SW curve of the $\calS_k$ theories knows about it and will illuminate the path leading to the symmetry algebra/representations underlying the 2D CFT. Following the $\calN=2$ class $\calS$ paradigm \cite{Alday:2009aq,Kozcaz:2010af,Keller:2011ek}, we first compare the meromorphic differentials  $\cuco_{\ell}$ of the SW curves derived in  \cite{Coman:2015bqq} with the weighted current correlation functions\footnote{ We define the $\vvac{J_{\ell} (t)}$ in section~\ref{subsec: Comparisons of the curves with the blocks}. For now, it suffices to point out that for the simplest case of three fields they can be computed as a ratio of correlation function $$\vvac{ J_{\ell} (t) }_3 = \frac{\vac{ J_{\ell} (t) \ver_1(x_1) \ver_2(x_2) \ver_3(x_3)}}{\vac{\ver_1(x_1) \ver_2(x_2) \ver_3(x_3)} },$$ with the $\ver_i$ being primary fields.} $\vvac{J_{\ell} (t)}$ computed on the CFT side. Specifically, the identification works in the semi-classical limit $\epsilon\rightarrow 0$
\beq
\label{eq:curvecurrentrelation}
\lim_{\epsilon\rightarrow 0} \vvac{ J_{\ell} (t) } = \cuco_{\ell}(t) \,  \nonumber
\eeq
where $\epsilon=\epsilon_{1}+\epsilon_2$, with the $\epsilon_i$ being the $\Omega$-background deformation parameters. Since the CFT primary fields enter in the computation of $\vvac{ J_{\ell} (t) }$, the above identification dictates to us their quantum numbers. In particular, we can learn the form of the CFT representations that the primary fields live in.

We discover that the spectral curves of the 4D  $\SU(N)$ gauge theories of class $\calS_{k}$ can be reproduced from the 2D CFT weighted current correlation functions of the $\bW_{Nk}$ algebra with \textit{non-unitary primary fields}.  This is based on the observation that the SW curves of $\SU(N)$  class $\calS_{k}$ theories can be obtained from the $\calN=2$ $\SU(Nk)$ curves by tuning the mass/Coulomb branch parameters appropriately. On the CFT side, one then simply computes  the conformal/W-blocks for $\bW_{Nk}$ with $Nk=2,3,4,\ldots$ and sets the parameters to appropriate values. In addition, we use the known AGT correspondence for the $\calN=2$ $\SU(Nk)$ theories to derive a conjecture for the $\calN=1$ class $\calS_k$ instanton partition functions.

\medskip

This article is structured as follows. We begin in section~\ref{sec:curves} by reviewing the construction of the SW curves for the class $\calS_k$ theories. We introduce some of their properties and discuss the weak coupling limit and the Gaiotto curve. The next section~\ref{sec: reviewAGT} is concerned with recapitulating some aspects of the AGT correspondence that are essential for our work such as the identifications of the parameters on both sides of the duality and the relationships between the 2D CFT blocks and the 4D instanton partition functions. Since this is a review section, the readers familiar with the AGT correspondence can move directly to the next section~\ref{sec: SkAGT correspondence} in which we present our main results concerning the structures of the CFT representations, the comparisons with the $\calS_k$ SW curves and the investigation of the (orbifold) Nekrasov instanton partition functions.  We conclude in section~\ref{app:conclusion and outlook} where we also overview some potential directions of future research that our article suggests. Most technical computations as well as bulky formulas are stored in the appendices. 
 
%!TEX root = ../WBlocks.tex
%%%%%%%%%%%%%%%%%%%%%%%%%%%%%%%%%%%%%%%%%%%%%

%----------------------------------------------------------------------------------------------
\section{The curves}
\label{sec:curves}
%----------------------------------------------------------------------------------------------

The starting point of our work is the SW curves. By comparing them to the 2D CFT 3 and 4-point blocks, we will discover the algebra and the representations that underly the 2D theory we are looking for.  In this section we present the SW curves and provide a short review of their derivation as well as of the important information they contain.

\begin{table}[h!]
\centering
\renewcommand{\arraystretch}{1.3}
\begin{tabular}{|c|c|c|c|c|c|c|c|c|c|c|c|}
\hline	
&$x^0$ & $x^1$ &$x^2$ &$x^3$ &$x^4$ &$x^5$ &$x^6$ &$x^7$ &$x^8$ &$x^9$ &($x^{10}$) \\
\hline
$M$ NS5 branes &$-$&$-$&$-$&$-$&$-$&$-$&.&.&.&.&.\\
\hline
$N$ D4-branes &$-$&$-$&$-$&$-$&.&.&$-$&.&.&.&$-$\\
\hline
$A_{k-1}$ orbifold&$.$&$.$&$.$&$.$&$-$&$-$&$.$&$-$&$-$&.&.\\
\hline
\end{tabular}
\caption{\it Type IIA brane configuration  for the 4D  $\mathcal{N}=1$ theories of
class $\mathcal{S}_k$.}
\label{configIIA}
\end{table}

\paragraph{Review of the type IIA/M-theory construction.} The class $\calS_k$ SW curves (with $k=1,2,\ldots$) were derived in \cite{Coman:2015bqq} following Witten's \cite{Witten:1997sc} M-theory construction in which the  implementation of the orbifold is very simple. The main points of it we outline here. The SW curves were originally introduced as auxiliary algebraic curves \cite{Seiberg:1994rs,Seiberg:1994aj}. Using type IIA string theory, $\mathcal{N}=2$ gauge theories can be realized as world volume theories on D4-branes, which are suspended between NS5-branes. Uplifting this brane setup to M-theory, all the branes can be seen as one single M5-brane with a non-trivial topology. The geometry of this M5-brane is encoded in the SW curve. Therefore, the SW curve can also be derived by studying the minimal surface of the M5-brane \cite{Witten:1997sc}.

\begin{figure}[htbp!]
             \begin{center}       
              \includegraphics[scale=0.25]{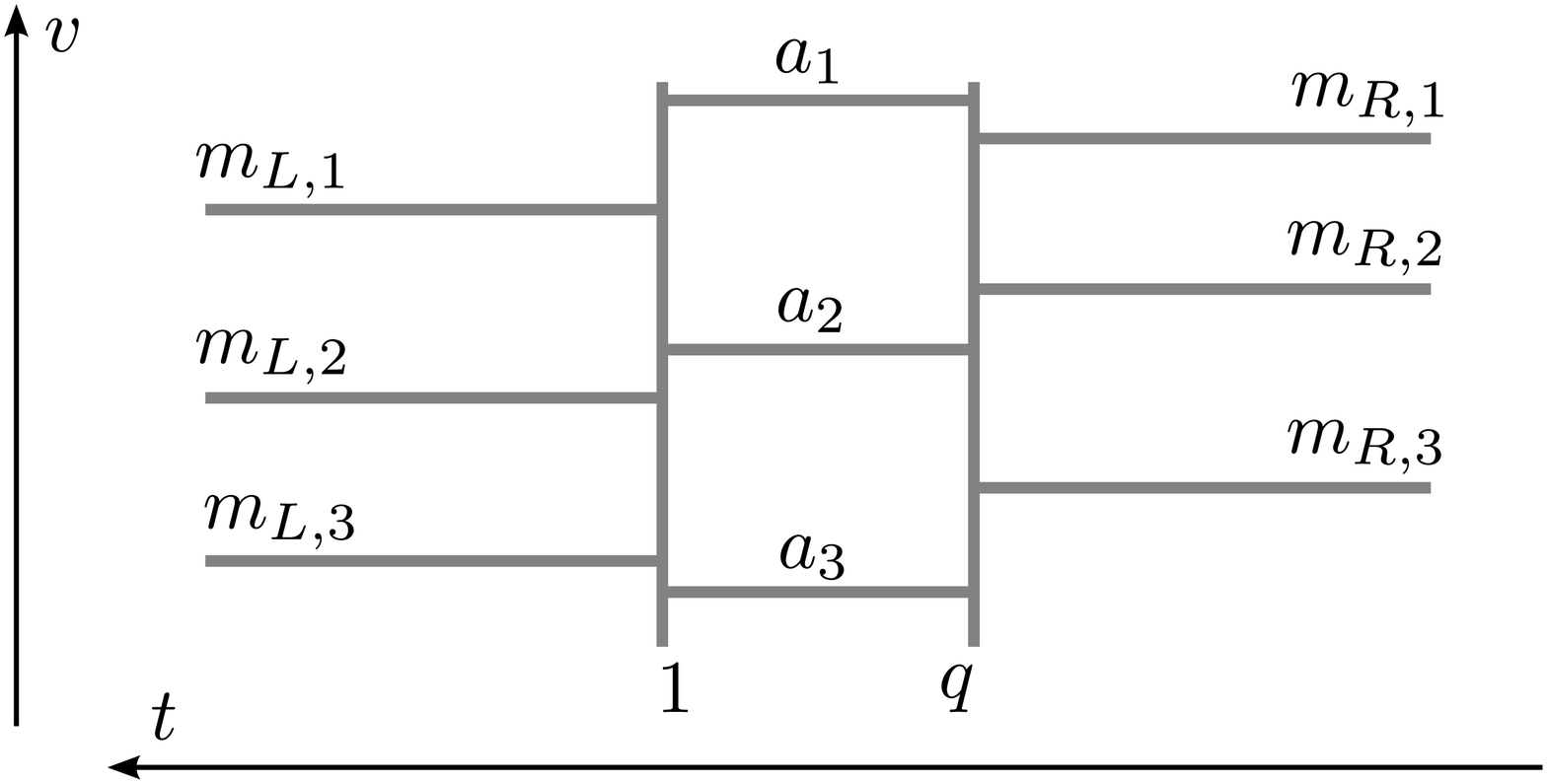}
              \caption{This figure illustrates the position of the branes (horizontal D4s and vertical NS5s) for the case of the $\calN=2$ $\SU(3)$ gauge theory. In the $\calN=1$ case, one needs to introduce an orbifold and image branes as reviewed in \cite{Coman:2015bqq}. From the equation for the curve \eqref{eq:curvewithgaugegroup}, we see that for $t\rightarrow 0/\infty$ the solutions of the curve are $v=\mL{i}/\mR{i}$, while for $v\rightarrow \infty$ the solutions are $t=1,q$.}
              \label{Fig:BranePosition}
            \end{center}
\end{figure}

The theories in  class  $\mathcal{S}_k$  can be realized through the type IIA string theory brane setup of table \ref{configIIA}, which was originally considered in \cite{Lykken:1997gy,Lykken:1997ub}. For $k=1$ there is no orbifold and one obtains the $\mathcal{N}=2$ theories of  class  $\mathcal{S}$  \cite{Gaiotto:2009we}. The $\SU(2)_R$ R-symmetry  of the $\mathcal{N}=2$ theories corresponds to the rotation symmetry of the coordinates $x^{7}$, $x^{8}$ and $x^{9}$ which is broken by the orbifold to the $\UU(1)_R$ symmetry of $x^{7}$, $x^{8}$ rotations. The rotation on the $x^{4}$, $x^{5}$ plane corresponds to the $\UU(1)_r$ symmetry of the $\mathcal{N}=2$ theories and is also lost \cite{Gaiotto:2009we}.
The SW curves are derived by uplifting IIA string theory to M-theory and they are functions of the holomorphic coordinates
\beq
  v \equiv x^4 + i x^5 ~, \qquad s \equiv x^6 + i x^{10}  \qquad \mbox{and} \qquad
 t\equiv e^{-\frac{s}{R_{10}}}  ~
\eeq
where $R_{10}$ is the M-theory circle. We follow the conventions of \cite{Bao:2011rc}. The orbifold action is imposed via the identification
\begin{equation}
\label{OrbifoldAction}
 v   ~     \sim     ~  e^{\frac{2\pi i}{k}}v  \, .
\end{equation}
The mass parameters  $\mL{i}$ and $\mR{i}$ are given by the asymptotic position of the M5 branes as $t\rightarrow 0$ and $t\rightarrow \infty$, while the coupling constant $q$ enters the setup via the asymptotic position of the M5 branes for $v \rightarrow \infty$, see figure~\ref{Fig:BranePosition} for an illustration.

\paragraph{The SCQCD curves.} The spectral curve that describes the Coulomb branch of the  $\mathbb{Z}_k$ orbifold daughter of $\mathcal{N}=2$ $\SU(N)$ SCQCD (SCQCD$_k$) is given by the equation
\beq
\label{eq:curvewithgaugegroup}
t^2\prod_{i=1}^N\left(v^k -\mL{i}^k \right) +t\left(-(1+q)v^{Nk}+\sum_{l=1}^Nu_{lk}v^{(N-l)k}\right)+ q\prod_{i=1}^N\left(v^k -\mR{i}^k \right) = 0\,.
\eeq
It is sometimes convenient to group the masses as $m_i=\mL{i}$ and $m_{N+i}=\mR{i}$ for $i=1,\ldots N$. We can rescale the variable $v$ as $v= xt$ and normalize the coefficient of the highest power in $x$ to one.\footnote{The Seiberg-Witten differential in these coordinates is given by $\lambda_{SW}=xdt$.} Thus, we can write the equation for the curve as
\beq
\label{eq:four point curve rescaled}
\sum_{\ell = 0}^{N} \cuco^{(4)}_{k\ell}(t) x^{k(N-\ell)}  =0\,,
\eeq
where the coefficients are given by $\cuco^{(4)}_{0}(t)=1$ and 
\beq
\label{eq:curvecoefficientsSCQCDgeneralNandk}
\cuco^{(4)}_{k\ell}(t)=\frac{(-1)^\ell \fc_L^{(\ell,k)}t^2+u_{ k\ell}t +(-1)^\ell \fc_R^{(\ell,k)}q }{t^{ k\ell}(t-1)(t-q)}\qquad \text{ for }\ell=1,\ldots, N\,.
\eeq
In the above, we have used the formula $\prod_{i=1}^N(v^k-m_i^k)=\sum_{s=0}^N(-1)^s\fc^{(s,k)}v^{k(N-s)}$ 
with the Casimirs (let use set for simplicity $\fc^{(s)}\equiv \fc^{(s,1)}$) defined as :
\beq
\label{eq:DefinitionCasimirs}
\fc^{(s,k)}=\sum_{i_1<\cdots < i_s=1}^Nm_{i_1}^k\cdots  m_{i_s}^k\,,\qquad \fc^{(0,k)}=1\,.
\eeq
For generic values of the masses, the Casimirs $\{\fc^{(s,k)}\}_{\ell=1}^N$ are algebraically independent of each other. 

We remark that one can perform an $\SL(2,\mathbb{Z})$ transformation $t\rightarrow \frac{a z +b}{c z +d }$, $x\rightarrow (cz+d)^2 x$ on the curve \eqref{eq:curvewithgaugegroup} and set $z_1=-\tfrac{d}{c}$, $z_2=-\tfrac{b-d}{a-c}$, $z_3=-\frac{b-d q}{a-c q}$ and  $z_4=-\tfrac{b}{a}$. This sends the singularities at $\infty$, $1$, $q$ and $0$ to the generic points $z_1$, $z_2$, $z_3$ and $z_4$ respectively. 

\begin{figure}[htbp!]
             \begin{center}       
              \includegraphics[scale=0.25]{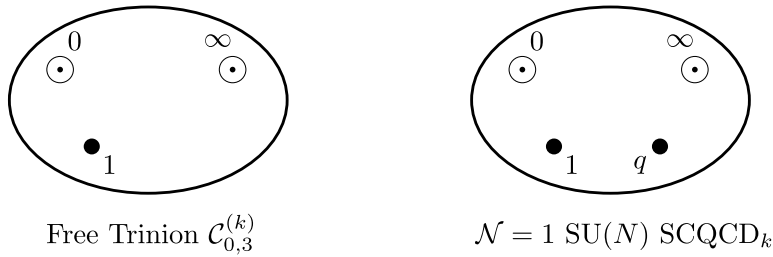}
              \caption{The UV curves of the trinion and of the SCQCD$_k$ theories. They are 3, respectively 4-punctured spheres. The full punctures are depicted by $\astrosun$ and placed at $t=0$ and $t=\infty$, while the simple punctures $\bullet$ are at $t=1$ and at $t=q$.}
              \label{Fig:UVcurves}
            \end{center}
\end{figure}

\paragraph{The free trinion curves.} As explained in \cite{Coman:2015bqq}, the free $\calC_{0,3}^{(k)}$ trinion curve can be obtained from the SCQCD$_k$ one by going to the weak coupling regime $q\rightarrow 0$ and identifying the Coulomb parameters $u_\ell$ appropriately with the masses. The resulting equation for the curve reads
\beq
\label{eq:curveTrinionGeneralOrbifold}
t \prod_{i=1}^N\left(v^k -\mL{i}^k \right) -\prod_{i=1}^N\left(v^k -\mR{i}^k \right) = 0\,.
\eeq
As before, we can rescale $v=xt$ and write the curve as $\sum_{\ell = 1}^{N} \cuco^{(3)}_{k\ell}(t) x^{k(N-\ell)}  =0$, 
with the curve coefficients (see \eqref{eq:DefinitionCasimirs} for the definition of the Casimirs) $\cuco^{(3)}_0=1$ and 
\beq
\label{eq:curvecoefficientstriniongeneralNandk}
\cuco^{(3)}_{k \ell}(t)=(-1)^{\ell}\frac{\fc_L^{(\ell,k)}t-\fc_R^{(\ell,k)}}{t^{k\ell}(t-1)}\quad \text{ for } \ell=1,\ldots, N\,.
\eeq
The above coefficients can be directly obtained by taking the limit $q\rightarrow 0$ in  \eqref{eq:curvecoefficientsSCQCDgeneralNandk} and setting 
\beq 
\label{eq:weak coupling limit of ukl}
u_{ k\ell }(q=0)\longrightarrow (-1) ^{\ell+1}\fc_R^{(\ell,k)}\,.
\eeq
The UV curves corresponding to the free trinion and to the SCQCD theories are depicted in figure~\ref{Fig:UVcurves}. They are three and respectively four punctured\footnote{The UV curves are characterized by the meromorphic differentials $\cuco^{(n)}_{s}$ that have only poles and no branch cuts.
The additional punctures $\bigstar$ discussed in \cite{Coman:2015bqq} will not be relevant for our purposes here.
} spheres with the punctures at $t=0$ and $t=\infty$ being full punctures $\astrosun$, while those at $t=1$ and $t=q$ are simple punctures $\bullet$, see \cite{Coman:2015bqq}. 

\paragraph{Gaiotto Shifts in $x$ for $k=1$.} Due to the orbifold relation \eqref{OrbifoldAction}, we are allowed to shift the variable $x$ for $k=1$, \textbf{but not} for $k>1$. This shift is the consequence of the additional $\UU(1)$ degrees of freedom  that are present for $k=1$ but, as we shall see more in detail later, disappear for $k>1$. 
For $k=1$, if we go from an equation $\sum_{i=0}^N x^i\cuco_i$ to $\sum_{i=0}^N x^i\cuco_i'$ by making the tranformation $x\rightarrow x-\kappa \cuco_1$, then we find
\beq
\label{eq:shiftinkappa}
\cuco_{\ell}'=\sum_{j=N-\ell}^N\binom{j}{N-\ell}\cuco_{N-j}(-\kappa \cuco_1)^{j+i-N}=\sum_{j=0}^{\ell}\binom{N-j}{N-\ell}\cuco_{j}(-\kappa \cuco_1)^{\ell-j}\,.
\eeq
We remind that $\cuco_0=1$ before and after the transformation. It is clear that the shift leaves the 2-form $\Omega_2=d\lambda_{SW}=dx\wedge dt$ unchanged, however the structure of the poles of $\lambda_{SW}$ on the various sheets of the curve does change, see \cite{Coman:2015bqq}. 
If we put the shift parameter $\kappa$ equal to $\tfrac{1}{N}$, then the coefficient $\cuco_1'$ vanishes - the resulting curve is known as the Gaiotto curve. Let us denote the curve coefficients for the Gaiotto curve by $\tilde{\cuco}_\ell^{(n)}$. As we shall review later, their expansion around the poles in $t$ gives the charges of the $\bW_N$ algebra. One easily computes
\beq
\label{eq:expansion of the curve coefficients for the simple punctures}
\begin{split}
&\tilde{\cuco}^{(3)}_\ell(t)=\frac{-\binom{N}{\ell}\frac{\ell-1}{N^\ell}\left(M_L-M_R\right)^\ell}{(t-1)^\ell}+\cdots\,, \\ &\tilde{\cuco}^{(4)}_\ell(t)=\frac{-\binom{N}{\ell}\frac{\ell-1}{N^\ell}M_L^\ell}{(t-1)^\ell}+\cdots\,,\qquad \tilde{\cuco}^{(4)}_\ell(t)=\frac{-\binom{N}{\ell}\frac{\ell-1}{N^\ell}\left(-M_R\right)^\ell}{(t-q)^\ell}+\cdots\,.
\end{split}
\eeq
In the above, we have introduced the left and right center of masses
\beq
M_L=\sum_{i=1}^N\mL{i}=\fc_L^{(1)}\,,\qquad M_R= \sum_{i=1}^N\mR{i}=\fc_R^{(1)}\,.
\eeq
It is useful to furthermore introduce the $\SU(N)$ masses
\beq
\label{eq:SUNmasses}
\mLt{i}= \mL{i}-\frac{M_L}{N}\,,\qquad \mRt{i}= \mR{i}-\frac{M_R}{N}\,,
\eeq
which obey $\sum_{i=1}^N\tilde{m}_i=0$. The corresponding Casimirs with $m\rightarrow \tilde{m}$ are denoted by $\tilde{\fc}^{(\ell)}$. 
Expanding the curve coefficients around $t=0$ and $t=\infty$ and using \eqref{eq:relationSUNandUNCasimirs}, we find that 
\beq
\tilde{\cuco}^{(n)}_\ell(t)=\frac{(-1)^{\ell}\tilde{\fc}_R^{(\ell)}}{t^\ell}+\cdots\,,
\eeq
for $n=3,4$. Performing the $\SL(2,\mathbb{Z})$ transformation is $t\rightarrow -\tfrac{1}{t}$, we can compute the expansion around $t=\infty$ and we get for $\tilde{\cuco}^{(n)}_\ell$ a pole of order $\ell$ with coefficient $\tilde{\fc}_L^{(\ell)}$.

%----------------------------------------------------------------------------------------------
\section{Review of some aspects of the AGT correspondence}
\label{sec: reviewAGT}
%----------------------------------------------------------------------------------------------

In this section, we wish to review the essentials of the AGT correspondence and especially of the elements that we shall need in the rest of the article. The essential elements are summarized in table~\ref{tab:AGToverview}. 

We begin with a short introduction of the Toda CFT and its symmetries. We then relate the charges of the Toda currents to the curves of the previous section and thus match the parameters. Following this, we explain how to recover the complete curve coefficients from the CFTs as ratios of conformal/W-blocks and relate the blocks to the instanton partition functions of the gauge theory.

%-------------------------------------------------------------------------------------
\begin{table}[ht!]
\centering
\renewcommand{\arraystretch}{1.6}
\begin{tabular}{|c|c|c|}

\hline
Gauge theory
& 
Toda  CFT
& 
Relations
\\
\hline
$\Omega$ deformation parameters $\epsilon_1$, $\epsilon_2$ & Coupling $b$  & $\epsilon_1=b$, $\epsilon_2=b^{-1}$, \\
$\epsilon\equiv \epsilon_1+\epsilon_2$ & $Q=b+b^{-1}$ &  $Q=\tfrac{\epsilon}{\sqrt{\epsilon_1\epsilon_2}}$
\\
Masses $m_i$ & Charges of the external states  $\ba_1, \ldots, \ba_n$  & \eqref{eq:Full punctures and masses}, \eqref{eq:TodachargesFullpunctures},\eqref{eq:Simple punctures and masses}, \eqref{eq:TodachargesSimplepunctures}
\\
Coulomb moduli $u_\ell$ & Charges of the intermediate states $\bw$ & \eqref{eq:relation for u_1}, \eqref{eq: u2 related to CFT data} \\
Coulomb branch parameters $\fa^{(\ell)}$  & Casimirs of the intermediate state $\ba$ \eqref{eq: charges of the intermediate state}  & \eqref{eq:DefinitionCasimirs for the Coulomb}\\
Full punctures $\astrosun$, see figure~\ref{Fig:UVcurves}& Primary fields $\ver_{\astrosun}$  \eqref{eq:defVbaforToda}, \eqref{eq:Full punctures and masses} & \\
Simple punctures $\bullet$, see figure~\ref{Fig:UVcurves} & Primary fields $\ver_{\bullet}$ \eqref{eq:defVbaforToda}, \eqref{eq:Simple punctures and masses}&\\
Shift $x\rightarrow x-\kappa\cuco_1$ in the curve \eqref{eq:shiftinkappa} & Redefinitions of the currents \eqref{eq:defcalJ}&\\
Instanton partition functions $\calZ^{\text{inst}}$ \eqref{eq:general Z inst}& W-blocks $\block$ \eqref{eq:definition of W - blocks}& \eqref{eq: Zinst and the blocks}\\
SW coefficients curve $\cuco^{(n)}_\ell$ \eqref{eq:four point curve rescaled}&  Ratios of W-blocks $\vvac{J}_n$ \eqref{eq:definition of vvac}& \eqref{eq:main curve CFT relation}, \eqref{eq:identification calJ with cuco}\\
$S^4$ partition function & Full correlation function& \eqref{eq:fullpartitionfunctionidentification}\\
\hline
\end{tabular}
\renewcommand{\arraystretch}{1.0}
\caption{ This table presents an overview of the elements of the AGT correspondence that we need as well as the equations where the identifications appear.}
\label{tab:AGToverview}
\end{table}

%----------------------------------------------------------------------------------------------
\subsection{The Toda CFT}
\label{subsec:TodaCFT}
%----------------------------------------------------------------------------------------------

We refer to the appendix B of \cite{Mitev:2014isa} for our conventions regarding the $\SU(N)$ weights $\hs_i$, simple roots $e_i$, fundamental weights $\omega_i$, Weyl vector $\rho$ and scalar product $\form{\cdot}{\cdot}$.

The action (see \cite{Fateev:2007ab}) of the $\SU(N)$ Toda theory in our normalizations reads (we define the $\bphi$ fields below)
\beq
\label{eq:The Toda action}
S_{\text{Toda}}=\int\left(\frac{1}{8\pi}\hat{g}^{mn}\form{\partial_{m}\bphi}{\partial_{n}\bphi}+\frac{\form{\fQ}{\bphi}}{4\pi}\hat{R}+\mu\sum_{j=1}^{N-1}e^{ b\form{e_j}{\bphi}}\right)\sqrt{\hat{g}}\,d^2x\,,
\eeq
where $\hat{g}_{mn}$ is the background metric and $\hat{R}$ is the corresponding scalar curvature coupling to the background charge $Q$. One defines $\fQ=Q\rho$ and relates $Q$ to the coupling $b$ via $Q=b+b^{-1}$ so that the theory is conformal. The cosmological constant $\mu$ is not particularly important and only enters the game through the overall normalization of the 3-point structure constants in the quantum theory. The central charge $c$ of the Toda CFT is given by  
\beq
c=N-1+12\form{\fQ}{\fQ}=(N-1)\left(1+N(N+1)Q^2\right)\,,
\eeq
so that $c=N-1$ for $Q=0$. We still have to explain the $N-1$ component field $\bphi$. In order to introduce some notation for later, we start (in the formal free case where the cosmological constant $\mu$ is zero) with the $N$ free fields $\{\varphi_j\}_{j=1}^N$ with the OPE $\varphi_i(z)\varphi_j(w)\sim -\delta_{ij}\log|z-w|^2$.  Next, we define the $\SU(N)$ field $\bphi$
\beq
\label{eq:def bphi and tilde phi}
\bphi=\sum_{j=1}^{N-1}\omega_j\tilde{\varphi}_j=\sum_{j=1}^{N-1}\omega_j(\varphi_j-\varphi_{j+1})=\sum_{i=1}^{N-1}\hs_i(\varphi_i-\varphi_N)\,,
\eeq
with $\omega_j$ being the $\SU(N)$ fundamental weights. The above implies that $\tilde{\varphi}_i(z)\tilde{\varphi}_j(w)\sim -\car_{ij}\log|z-w|^2$, where $\car_{ij}=2$ if $i=j$, $-1$ if $|i-j|=1$ and zero otherwise is the $\SU(N)$ Cartan matrix. The $\UU(1)$ free field that decouples from the rest of the Toda action is $\fb=\frac{1}{\sqrt{N}}\sum_{j=1}^N\varphi_j$ with the free field OPE $\fb(z)\fb(w)\sim -\log|z-w|^2$.  The original $\varphi_j$ fields can be written as $ \varphi_j=\frac{1}{\sqrt{N}}\lambda+\form{\hs_j}{\bphi}$.
Using the field $\bphi$ in the free limit is straightforward since we have the OPE
\beq
\label{eq:bphiWickcontraction}
(\balpha,\bphi)(z)(\bbeta,\bphi)(w)\sim -(\balpha,\bbeta) \log|z-w|^2\,,
\eeq
which follows from the identity \eqref{eq:summationFormula2}.

The quantum Miura transform (see for example \cite{Fateev:2011hq, Kanno:2009ga}) relates the currents of the $\bW_N$ algebra for the Toda theory in terms of the $N-1$ free fields $\bphi$. One roughly speaking sets $\mu=0$ in \eqref{eq:The Toda action} and expands the Lax operator  $\widetilde{\mathcal{R}}_N$ as
\beq
\label{eq:Laxoperator}
\widetilde{\mathcal{R}}_N=:\prod_{j=1}^{N}(Q\partial_z + (\hs_{j},\partial\bphi(z))):=\sum_{s=0}^N\calW_{s}(z)(Q\partial_z)^{N-s}\,,
\eeq
where $::$ denotes normal-ordering. 
Note that the $\calW_s$ coming from the quantum Miura transform are for $s>2$ in general \textbf{not} conformal primaries. They differ from the $\bW_N$ currents $W_s$ by terms proportional to $Q$ and hence agree (up to a convention dependent normalization that for us is set to one) for $Q=0$. 
We remind that the OPEs of the $\bW_N$ currents with a primary field $\ver_{\ba}$ are
\beq
\label{eq: action of W_s on a primary}
W_s(z_1)\ver_{\ba}(z_2,\bar z_2)\sim \frac{\charge_s(\ba)}{(z_1-z_2)^s}\ver_{\ba}(z_2,\bar z_2)+\sum_{n=1}^{s-1}\frac{W_{s,-n}\ver_{\ba}(z_2,\bar z_2)}{(z_1-z_2)^{s-n}}\,.
\eeq
Here the $W_{s,-n}$ denote  the lowering modes of the $W_s$ current. We parametrize the primary fields/ vertex operators in terms of $\SU(N)$ weights $\ba$ as\footnote{The primary fields $\ver$ also carry a $\fb$ dependent part as we write later in \eqref{eq:complete vertex operators with fb}, but we can ignore that part for now.}
\beq
\label{eq:defVbaforToda}
\ver_{\ba}(z)=e^{(\ba,\bphi)}(z)\,.
\eeq
From this parametrization of the primary fields, using $(\hs_j,\partial \bphi)(z)\ver_{\ba}(w) \sim -\frac{\form{\hs_j}{\ba}}{z-w}\ver_{\ba}(w)$ as well as the general relation ($u$ and $d_j$ are arbitrary complex parameters)
\beq
\prod_{j=1}^N\left(u\partial_z+\frac{d_j}{z}\right)=\sum_{s=0}^N\frac{1}{z^s}\left[\sum_{i_1<i_2<\cdots< i_s=1}^N\prod_{m=1}^s\big(d_{i_m}-u(k-m)\big)\right] (u\partial_z)^{N-s}\,,
\eeq
we derive the charges of the $\calW_{s,0}$ modes to be (see also \cite{Fateev:1987zh})
\beq
\label{eq:defConformalDimensionsforToda}
\begin{split}
\cd(\ba)&=\charge'_2(\ba)=\sum_{i<j=1}^N\form{\hs_i}{\ba}\form{\hs_j}{\ba}+Q\sum_{j=2}^N(j-1)\form{\hs_j}{\ba}=\frac{\form{2\fQ-\ba}{\ba}}{2}\,,\\
\charge'_s(\ba)&=(-1)^s\sum_{i_1<i_2<\cdots< i_s=1}^N\prod_{j=1}^s\left(\form{\hs_{i_j}}{\ba}+Q(s-j)\right)\,,
\end{split}
\eeq
where we have used \eqref{eq:summationFormula2} and $\fQ= Q \rho$ with $\rho=\sum_{j=2}^N(j-1)\hs_j$. The charges of the primary $\bW_s$ fields with modes $W_{s,0}$ with $s>2$ differ from the above. For example $\charge_3(\ba)=\charge_3'(\ba)+Q(N-2)\charge_2'(\ba)$, which can be  rewritten as 
\beq
\label{eq:chargeW3 for general Q}
\charge_3(\ba)=-\sum_{i_1<i_2<i_3=1}^3\prod_{s=1}^3\form{\ba-\fQ}{\hs_{i_s}}\,,
\eeq
see also \cite{Bouwknegt:1992wg} for more details.

The  limit $Q\rightarrow 0$ is referred to as the ``semi-classical'' limit\footnote{This is different from the semi-classical limit $b\rightarrow \infty$ of the Toda CFT considered for example in \cite{Fateev:2007ab}. } and it is defined by the substitution $Q\partial_z\longrightarrow x$ in \eqref{eq:Laxoperator}. 
This limit is called semi-classical because it replaces the pair $(Q \partial_z, z)$ that satisfies the Heisenberg commutation relations with the commuting variables $(x,z)$. In that limit, we have $W_s=\calW_s$ and hence
\beq
\label{eq:Wncurrents}
\begin{split}
W_{s}&=\sum_{1\leq j_1<j_2<\cdots<j_s\leq N}(\hs_{j_1},\partial \bphi)\cdots (\hs_{j_s},\partial \bphi)\quad \Longrightarrow\quad  T=-\frac{\form{\partial \bphi}{\partial \bphi}}{2}\,,\\
\lim_{Q\rightarrow 0}\charge_s(\ba)&=\lim_{Q\rightarrow 0}\charge'_s(\ba)=(-1)^s\sum_{1\leq j_1<j_2<\cdots<j_s\leq N}(\hs_{j_1}, \ba)\cdots (\hs_{j_s}, \ba)\,.
\end{split}
\eeq
One of the consequences of the AGT correspondence is that the semi-classical limit of the Lax operator reproduces the Seiberg-Witten curve after an $x$ shift to the Gaiotto curve
\beq
\label{eq: curve to block comparison Main}
\vvac{\widetilde{\mathcal{R}}(x)}=\sum_{\ell=0}^N x^{N-\ell}\vvac{W_s(t)}=\sum_{\ell=0}^Nx^{N-\ell}\tilde{\cuco}_{\ell}(t)=0\,,
\eeq
since as we shall review in section~\ref{subsec: Comparisons of the curves with the blocks}, 
\beq
\label{eq:main curve CFT relation}
\lim_{Q\rightarrow 0}\vvac{W_s(t)}=\tilde{\cuco}_{\ell}(t)\,.
\eeq
We refer to \eqref{eq:shiftinkappa} and its surrounding paragraph for the definition of the curve coefficients $\tilde{\cuco}_{\ell}(t)$.
We shall also see that \eqref{eq: curve to block comparison Main} can be made to work also for the case without the shift in $x$. This requires the reintroduction of the decoupled $\UU(1)$ degrees of freedom that on the CFT side are contained in the free boson field $\fb$.

%----------------------------------------------------------------------------------------------
\subsection{Identification of the parameters}
\label{subsec:parameter identification}
%----------------------------------------------------------------------------------------------

\begin{figure}[tbp!]
             \begin{center}       
              \includegraphics[scale=0.25]{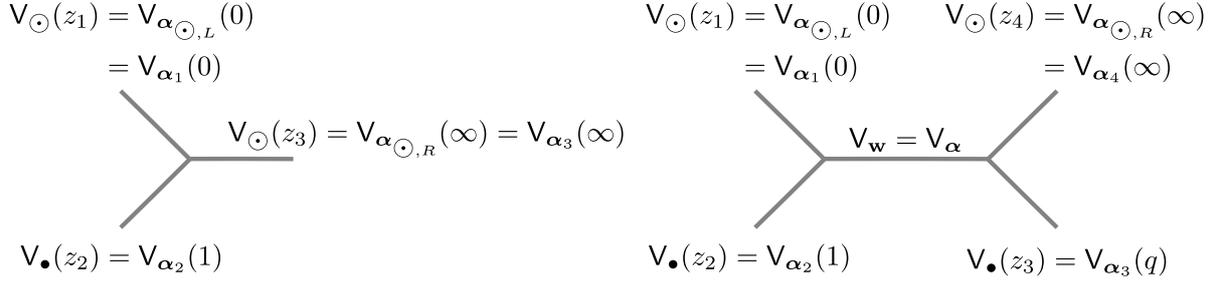}
              \caption{This figure illustrates the parametrization of the primary fields of the Toda CFT for the 3 and 4-point case. It indicates in particular which fields are full and which are simple punctures.}
              \label{Fig:34pointParametrization}
            \end{center}
\end{figure}

  In order to make \eqref{eq:main curve CFT relation} precise, we need to first relate the Toda CFT charges $\ba$ of the primary fields \eqref{eq:defVbaforToda} with the mass and Coulomb parameters appearing in the curves.  From the curves, we have $2N$ mass parameter $\mL{i}$ and $\mR{i}$ with $i=1,\ldots, N$ and we defined in \eqref{eq:SUNmasses} the $\SU(N)$ masses $\mLt{i}$ and $\mRt{i}$ as well as the centers of mass $M_L$ and $M_R$.
The masses are related to the weights $\balpha_{\astrosun}$ of the full punctures $\ver_{\astrosun}$ via
\begin{align}
\label{eq:Full punctures and masses}
&\mLt{i}= -\form{\balpha_{\astrosun, L}-\fQ}{\hs_i}\,,& &\mRt{i}= \form{\balpha_{\astrosun, R}-\fQ}{\hs_i}\,,&\nonumber\\
&\balpha_{\astrosun, L}=\sum_{i=1}^{N-1}(-\mLt{i}+\mLt{i+1}+Q)\omega_i\,,& &\balpha_{\astrosun, R}=\sum_{i=1}^{N-1}(\mRt{i}-\mRt{i+1}+Q)\omega_i\,.&
\end{align}
Thus, for the case of three points, $\mLt{i}= -\form{\balpha_{1}-\fQ}{\hs_i}$ and $\mRt{i}= \form{\balpha_{3}-\fQ}{\hs_i}$, while for the case of four points the parametrization becomes $\mLt{i}= -\form{\balpha_{1}-\fQ}{\hs_i}$ and $\mRt{i}= \form{\balpha_{4}-\fQ}{\hs_i}$. Equation \eqref{eq:Full punctures and masses} and \eqref{eq:defConformalDimensionsforToda} imply for $Q=0$ that the $W_s$ charges of the full punctures $\ver_{\astrosun}$ are equal to 
\beq
\label{eq:TodachargesFullpunctures}
\charge_s(\ba_{\astrosun,L})=\tilde{\fc}^{(s)}_L\,,\qquad \charge_s(\ba_{\astrosun,R})=(-1)^s\tilde{\fc}^{(s)}_R\,.
\eeq
On the other hand, the weights of the simple punctures $\ver_\bullet$ are given by 
\beq
\label{eq:Simple punctures and masses}
\ba_{\bullet}=-\varkappa \omega_{N-1}\,,
\eeq
where $\varkappa$ depends on the number of punctures. 
 For the three points case $\ba_2=-(M_L-M_R) \omega_{N-1}$, while for the four point case $\ba_2=-(M_L-\fa^{(1)})\omega_{N-1}$ and $\ba_3=(M_R-\fa^{(1)})\omega_{N-1}$. The Casimir $\fa^{(1)}=\sum_{i=1}^Na_i$ comes from the intermediate field in the 4-point block, see \eqref{eq:DefinitionCasimirs for the Coulomb} below, as well as figure~\ref{Fig:BranePosition}. The parametrization of the primary fields is also summarized in figure~\ref{Fig:34pointParametrization}.
It follows from \eqref{eq:Simple punctures and masses} that the corresponding $\bW_N$ charges for $Q=0$ are given by 
\beq
\label{eq:TodachargesSimplepunctures}
\begin{split}
\charge_s(\ba_\bullet)&=\varkappa^s\sum_{i_1<\cdots <i_s=1}^N\form{\omega_{N-1}}{\hs_{i_1}}\cdots \form{\omega_{N-1}}{\hs_{i_s}}=\varkappa^s\left(\sum_{i_1<\cdots <i_s=1}^{N-1}\frac{1}{N^s}+\sum_{i_1<\cdots <i_{s-1}=1}^{N-1}\frac{1}{N^{s-1}}\frac{1-N}{N}\right)\\
&=\frac{\varkappa^s}{N^s}\binom{N-1}{s}\left(1+\frac{(1-N)s}{N-s}\right)=-\binom{N}{s}\frac{(s-1)\varkappa^s}{N^s}\,,
\end{split}
\eeq
where we have used $\form{\omega_{N-1}}{\hs_j}=\tfrac{1}{N}$ for $j<N$ and $\form{\omega_{N-1}}{\hs_N}=\tfrac{1-N}{N}$. 

The last parametrization that we need to discuss is that of the Coulomb moduli $u_\ell$ of the curves that are related to the intermediate state $\ba$ in the 4-point block introduced in the next section~\ref{subsec: blocks and the instanton partition functions}, see also figure~\ref{Fig:34pointParametrization}. Similarly to the case of the full punctures \eqref{eq:Full punctures and masses}, we put 
\beq
\label{eq: charges of the intermediate state}
\ba=\sum_{i=1}^{N-1}(a_i-a_{i+1}+Q)\omega_{i}\quad \Longleftrightarrow \quad a_i= \form{\balpha-\fQ}{\hs_i}\,.
\eeq
It is useful to define the Casimirs for the parameters $a_i$ as in \eqref{eq:DefinitionCasimirs}, i.e. 
\beq
\label{eq:DefinitionCasimirs for the Coulomb}
\fa^{(s,k)}=\sum_{i_1<\cdots < i_s=1}^Na_{i_1}^k\cdots  a_{i_s}^k\,,\qquad \fa^{(0,k)}=1\,,
\eeq
where again $\fa^{(s)}\equiv \fa^{(s,1)}$. 
As we shall see in section~\ref{subsec: Comparisons of the curves with the blocks}, the Coulomb moduli $u_\ell$ are expressed via the Casimirs $\fc_L$, $\fc_R$ and \eqref{eq:DefinitionCasimirs for the Coulomb}. We also define for $k=1$ the Casimirs $\tilde{\fa}^{(s)}$ obtained by applying the definition \eqref{eq:DefinitionCasimirs for the Coulomb} to the $\tilde{a}_i=a_i-\frac{1}{N}\sum_{j=1}^Na_j$. From \eqref{eq:TodachargesFullpunctures} we see that for $Q=0$ the $\bW_N$ charges of the intermediate state are $\charge_s(\ba)=(-1)^s\tilde{\fa}^{(s)}$.

%----------------------------------------------------------------------------------------------
\subsection{The W-blocks and the instanton partition functions}
\label{subsec: blocks and the instanton partition functions}
%----------------------------------------------------------------------------------------------

\paragraph{Overview of the blocks. }In any CFT, knowledge of the correlation functions of two (i.e.~of the conformal dimensions $\cd$) and three point functions (i.e.~of the structure constants $C_{ijk}$) completely determines the higher point functions. For ordinary CFTs, it is enough to know the three-point functions of the Virasoro primary fields - the ones involving descendant field being then automatically determined. On the other hand, $\bW_N$ symmetry for $N>2$, while stronger than Virasoro, is not sufficient to determine the correlation functions of all descendant fields just from the knowledge of the correlation functions of the $\bW_N$ primaries.  Thankfully, for the cases that we consider here, some of the primary fields are short which imposes a sufficient number of extra conditions allowing for the derivation of the 3-point functions and then of the W-blocks.

Once the 2 and 3-point functions are known, the $n$-point functions can be determined by expanding in conformal/ W-blocks (see for example \cite{Mironov:2009dr} for a review). The blocks $\block$ are purely kinematic/symmetry quantities that are theory independent - they depend only on the charges $\bcharge$ of the fields (both the external $n$ ones as well as the intermediate ones) on the positions $q_1,\ldots, q_{n-3}$ that are not fixed by conformal symmetry and on the central charge $c$. The whole theory dependent information is contained in the 3-point structure constants $C_{ijk}$. 

Let us review  the 4-point $\bW_2$ case of Liouville theory for simplicity. Putting the points $z_1,\ldots, z_4$ to $\infty, 1, q, 0$ respectively, the full 4-point correlation function\footnote{Recall that the AGT correspondence identifies that full correlation function with the $S^4$ partition function: 
\beq 
\label{eq:fullpartitionfunctionidentification} 
\vac{\ver_1(\infty)\ver_2(1)\ver_3(q,\bar q)\ver_4(0)}\propto \calZ^{S^4}\,,
\eeq  
where the proportionality constant is not important here.
For the correlation function \eqref{eq:blockdecomposition of the 4 point function}, it is the partition function  of the $\SU(2)$ SCQCD theory with $N_F=4$.} can be expanded (in the $s$-channel) as
\beq
\label{eq:blockdecomposition of the 4 point function}
\vac{\ver_1(\infty)\ver_2(1)\ver_3(q,\bar q)\ver_4(0)}=\int d\alpha (C_{12\alpha}H_{\alpha\alpha}^{-1}C_{\alpha 34})\left|q^{\cd_\alpha-\cd_3-\cd_4}\block_{\cd_\alpha}(\cd_1,\cd_2,\cd_3,\cd_4|q)\right|^2\,,
\eeq
where $\alpha$ labels\footnote{For $N=2$ one sets $\ba=2\alpha\omega_1$. In general, the physical Toda fields obey $\text{Re}(\ba)=\fQ$. } the intermediate state in the OPE decomposition, and the integral is done over the space of physical Virasoro fields: 
$ \alpha\in \frac{Q}{2}+i \mathbb{R}$ with $\cd_\alpha=\alpha(Q-\alpha)$. 
The $H_{\alpha\beta}=\bracket{\ver_\alpha}{\ver_\beta}$ is an orthonormalization constant that is zero if $\alpha\neq \beta$ and that can be absorbed in the normalization of the primary fieds. 

Having introduced the decomposition of the full 4-point correlation function in terms of blocks in the Liouville case, we now want to concentrate on the blocks $\block$ and to consider them for the general $\bW_N$ case. They can be expanded in a power series in $q$ as 
\beq
\label{eq:definition of W - blocks}
\block_\bw(\bcharge_1,\bcharge_2,\bcharge_3,\bcharge_4|q)=\sum_{\bY,\bY',|\bY|=|\bY'|}q^{|\bY|}\gamma_{12\bw}(\bY)\shap_{\bw}^{-1}(\bY,\bY')\gvt_{\bw;34}(\bY')\,.
\eeq
In order to understand the above, we need to introduce all the ingredients (namely the charges $\bw$, the 3-point blocks/vertices $\gamma_{12\bw}$ and $\gvt_{\bw;34}$ as well as the Shapovalov form $\shap_{\bw}$) which requires some work. We start by reminding that the currents of the $\bW_N$ algebra are the $\{W_s(z)\}_{s=2}^N$. The currents are expanded in modes as
$W_{s}(z)=\sum_{n=-\infty}^\infty z^{-n-s}W_{s,n}$. We often write $L_n\equiv W_{2,n}$ as well as sometimes $W_n\equiv W_{3,n}$ if confusion can be avoided. Then we can straightforwardly define the elements needed for the blocks \eqref{eq:definition of W - blocks}:

\begin{itemize}
\item A highest weight Verma module  of the $\bW_N$ algebra is spanned by the vectors $W_{-\bY}\ver_\bw$, where 
\beq
\bw\define\{\cd,\charge_3,\charge_4,\ldots, \charge_N\}
\eeq
 are the $\ver_\bw$ charges of the $W_{n,0}$ generators and $\ver_\bw$ is annihilated by all the positive mode generators. We use the symbol $\ver_\bw$ both for the state in the Hilbert space and for the vertex operator that corresponds to it.
The descendant states are labeled by a set $\bY=\{Y_2;Y_3,\ldots, Y_N\}$ with each $Y_s=\{Y_{s,1},Y_{s,2},\ldots \}$ a partition of integers (arranged as $Y_{s,i}\leq Y_{s,i+1}$). The state $W_{-\bY}\ver_\bw$ is explicitly written as
\beq
W_{-\bY}\ver_\bw=\left(W_{2,-Y_{2,1}}W_{2,-Y_{2,2}}\cdots \right)\left(W_{3,-Y_{3,1}}W_{3,-Y_{3,2}}\cdots \right)\cdots \left(W_{N,-Y_{N,1}}W_{N,-Y_{N,2}}\cdots \right)\ver_\bw\,.
\eeq
For example, for $N=3$, $W_{-\{\{1,1,2\};\{2\}\}}\ver_\bw=L_{-1}^2L_{-2}W_{-2}\ver_\bw$. The conformal dimension of the state $W_{-\bY}\ver_\bw$ is equal to $\cd+|\bY|$ with $|\bY|=\sum_{s=2}^N|Y_s|$. The action of the other zero modes $W_{s,0}$ on the descendant states is in general not diagonal. 
\item The Shapovalov form $\shap$ is the scalar product of vectors in the Verma module
\beq
\shap_{\bw}(\bY,\bY')=\bracket{W_{-\bY}\ver_\bw}{W_{-\bY'}\ver_\bw}\,,
\eeq
where we demand that the scalar product obeys $\bracket{W_{s,-n}\ver_1}{\ver_2}=\bracket{\ver_1}{W_{s,n}\ver_2}$.
\item An important object is the 3-point W-block/vertex $\gv_{12\bw}(\bY)$. For our purposes, it is defined as the ratio of a 3-point function of two primary fields and one descendant $W_{-\bY}\ver_\bw$ to the 3-point function of just the primary fields:
\beq
\gv_{12\bw}(\bY)=\frac{\vac{\ver_1(\infty)\ver_2(1)\left(W_{-\bY}\ver_{\bw}\right)(0)}}{\vac{\ver_1(\infty)\ver_2(1)\ver_\bw(0)}}\,.
\eeq
Of course, it is possible to consider the cases in which $\ver_1$ or $\ver_2$ are not primary, but we do not need them here.
\item A similar object to $\gv$ is the vertex 
\beq
\gvt_{\bw;34}(\bY)=\frac{\bracket{W_{-\bY}\ver_\bw}{\ver_3(1)\ver_4(0)}}{\bracket{\ver_\bw}{\ver_3(1)\ver_4(0)}}\,,
\eeq
i.e.~the normalized scalar product of a state with the product of two primary fields inserted at $1$ and at $0$.
While for the Virasoro case, there is no need to introduce the $\gvt$ since $\gvt_{\cd;34}=\gv_{43\cd}$ (see the recursion relations \eqref{eq:iterativeequationsforgamma}), this is not true anymore for the general $\bW_N$ algebra.
\end{itemize}

\begin{figure}[htbp!]
             \begin{center}       
              \includegraphics[scale=0.25]{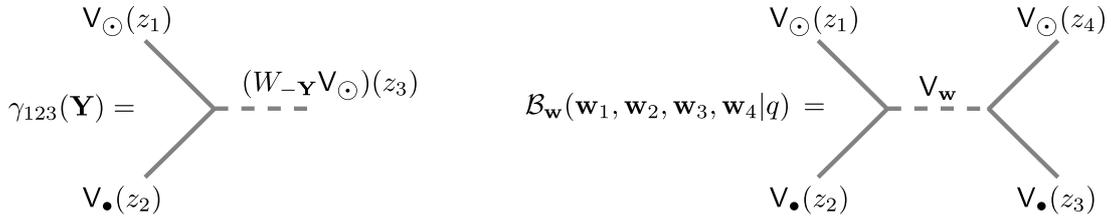}
              \caption{This figure depicts the three and four point W-blocks. Using conformal symmetry, for three points, we set $z_1=\infty$, $z_2=1$ and $z_3=0$, while for four points, we put $z_1=\infty$, $z_2=1$, $z_3=q$ and $z_4=0$. The dashed lines indicate descendant fields.}
              \label{Fig:34points}
            \end{center}
\end{figure}

One can depict the 3 and 4-point blocks graphically as sketched in~\ref{Fig:34points}.

\paragraph{The instanton partition functions and the blocks. }  The AGT correspondence identifies the Nekrasov instanton partition function $\calZ_{\text{inst}}$ to the W-blocks, after an appropriate factor has been removed. In the case that we are dealing with, namely for the $\calN=2$ $\SU(N)$ SCQCD with $N_F=2N$, the instanton partition function reads
\beq
\label{eq:general Z inst}
\begin{split}
\calZ_{\text{inst}}&=\sum_{\bY}q^{|\bY|}\calZ_{\text{vec}}(\textbf{a},\bY)\prod_{i=1}^N\calZ_{\text{antifund}}(\textbf{a},\bY;-\mL{i})\prod_{j=1}^N\calZ_{\text{fund}}(\textbf{a},\bY;\mR{j})\,,
\end{split}
\eeq
where $\textbf{a}=(a_1,\ldots, a_N)$  and $\bY=\{Y_1,\ldots, Y_N\}$ is a set of $N$ Young diagrams and the building blocks of $\calZ_{\text{inst}}$ are defined in appendix~\ref{app: Instanton Partition Functions}.
The partition function is related to the W-blocks as 
\beq
\label{eq: Zinst and the blocks}
\calZ_{\text{inst}}=\block_{\UU(1)}\block_{\bw}(\bw_1,\bw_2,\bw_3,\bw_4|q)\,.
\eeq
We remark that to relate the CFT data to the 4D Nekrasov partition functions, one should rescale all parameters with dimension of mass as $m\rightarrow \tfrac{m}{\sqrt{\epsilon_1\epsilon_2}}$ and also replace $Q\rightarrow \tfrac{\epsilon}{\sqrt{\epsilon_1\epsilon_2}}$.

The $\bW_N$ algebra charges $\bw_i$ are obtained by using the parametrization for $\ba_i$ in section~\ref{subsec:parameter identification} and using the identities \eqref{eq:defConformalDimensionsforToda}, \eqref{eq:chargeW3 for general Q}. 
 The $\UU(1)$ contribution, the 4-point block $\block_{\UU(1)}$, is given by the formula \eqref{eq:U1blockgeneraldefinition} derived in  appendix~\ref{app:U1blocks}
\beq
\label{eq:NekrasovConformalBlockU1}
\block_{\UU(1)}=(1-q)^{p_2p_3}=(1-q)^{\frac{(M_L-\fa^{(1)}) (M_R-\fa^{(1)}-N \epsilon)}{N \epsilon_1 \epsilon_2}}
\eeq
with the charges $p_2=-i\frac{M_L-\fa^{(1)}}{\sqrt{N \epsilon_1 \epsilon_2}}$ and $p_3=i\frac{M_R-\fa^{(1)}-N\epsilon}{\sqrt{N \epsilon_1 \epsilon_2}}$ (compare with \eqref{eq:definitionpiforN1U1}). In the above, we have used $\sum_{i=1}^Na_i=\fa^{(1)}$, see \eqref{eq:DefinitionCasimirs for the Coulomb}.

%----------------------------------------------------------------------------------------------
\subsection{Comparisons of the curves with the blocks}
\label{subsec: Comparisons of the curves with the blocks}
%----------------------------------------------------------------------------------------------

We now want to compare the curve coefficients $\cuco_\ell$ with the $\bW_N$ blocks, for three and for four points. In order to connect the blocks with the curve, we need to introduce yet another object, namely the \textit{3-point W-block with the insertion of an arbitrary current $J(t)$ at point $t$}. We write it as 
\beq
\label{eq:definition gvJ}
\gvJ_{12\bw}(J(t);\bY)\define\frac{\vac{\ver_1(\infty)\ver_2(1)J(t)\left(W_{-\bY}\ver_{\bw}\right)(0)}}{\vac{\ver_1(\infty)\ver_2(1)\ver_\bw(0)}}\,.
\eeq
The numerator of the above quantity is strictly speaking a 4-point function, but since $J(t)$ is a symmetry current and not an arbitrary object, the dependence of $t$ can be obtained by expanding $J(t)$ in modes and using the blocks $\gv_{12\bw}(\bY)$. Thus, we refer to $\gvJ_{12\bw}(J(t);\bY)$  as a 3-point block with an insertion of a current.

Armed with that definition, we define the weighted current correlation functions $\vvac{J(t)}$ as  the following ratio of blocks: 
\beq
\label{eq:definition of vvac}
\vvac{J(t)}_n\define\frac{\text{$n$-point W-block with insertion of $J(t)$}}{\text{$n$-point W-block}}\,,
\eeq
where the $n$-point W-block are computed with for $n$ primary fields. In the cases that concern us, two of the primary fields are full punctures $\ver_{\astrosun}$ placed at $z_1$ and $z_n$ and the remainig $n-2$ ones are simple punctures $\ver_{\bullet}$ at the points $z_2,\ldots, z_{n-1}$. By a conformal transformation, we place $z_1=\infty$, $z_2=1$ and $z_n=0$. In particular, for three points, we have for three primary fields
\beq
\vvac{J(t)}_3=\frac{\gvJ_{123}(J(t);\emptyset)}{\gv_{123}(\emptyset)}=\gvJ_{123}(J(t);\emptyset)=\frac{\vac{\ver_1(\infty)\ver_2(1)J(t)\ver_{\bw}(0)}}{\vac{\ver_1(\infty)\ver_2(1)\ver_\bw(0)}}\,.
\eeq
For four points, we have to specify the representation flowing in the middle with the label $\bw$. Labeling the point $z_3$ by $q$, the quantity $\vvac{J(t)}_4$ can be written as a power series expansion in $q$ as 
\beq
\label{eq:general4ptvvacFormula}
\vvac{J(t)}_4=\frac{\sum_{\bY,\bY',|\bY|=|\bY'|}q^{|\bY|}\gvJ_{12\bw}(J(t);\bY)\shap_{\bw}^{-1}(\bY,\bY')\gvt_{\bw;34}(\bY')}{\sum_{\bY,\bY',|\bY|=|\bY'|}q^{|\bY|}\gamma_{12\bw}(\bY)\shap_{\bw}^{-1}(\bY,\bY')\gvt_{\bw;34}(\bY')}\,.
\eeq
We note that in the above, if $J(t)$ is a spin $s$ current, the sum over the partitions $\bY=\{Y_2,\ldots, Y_N\}$ contains only those $\bY$ with $Y_{s+1}=\cdots = Y_{N}=\emptyset$.

 We now want to illustrate how the $\vvac{J_s}_n$ reproduce (see \eqref{eq:curvecurrentrelation}) the curve coefficients $\cuco_s^{(n)}$ for a few select cases. The comparisons with the curve coefficients in \textit{the rest of this section are all done in the limit $Q\rightarrow 0$}.

\paragraph{The $\UU(1)$ current. } Before we can make \eqref{eq:curvecurrentrelation} precise, we need to discuss how the $\UU(1)$ degrees of freedom contained in the free boson $\fb$, defined in section~\ref{subsec:TodaCFT}, affect the identification. 
For $k=1$, i.e. for the $\calN=2$ theories, we are allowed to shift $x\rightarrow x-\kappa \cuco_1$ in the curve. The Gaiotto curve with coefficients $\tilde{\cuco}_s$ is obtained for $\kappa=\tfrac{1}{N}$ and for that curve we have the identification \eqref{eq: curve to block comparison Main} between the ratios of blocks with insertions of the Toda currents $W_s$ and the curve coefficients $\tilde{\cuco}_s$. We can of course now perform the inverse shift $x\rightarrow x+\tfrac{1}{N} \cuco_1$. One might then ask how the currents should be modified in order for the ratio of blocks to give $\cuco_s$. The answer lies in bringing back to the game the free boson $\fb$. We define $J_1=i\partial \fb$ be the spin 1 free boson current.  We demand that in our normalizations 
\beq
\label{eq:J1 and cuco1}
\vvac{J_1(t)}_n\stackrel!=-i\sqrt{N}\cuco_1^{(n)}(t)\,.
\eeq
Since $\fb$ is completely decoupled from the Toda action, we can simply shift $x\rightarrow x-i\frac{1}{\sqrt{N}}J_1$ in \eqref{eq: curve to block comparison Main} and get for the Lax operator (remember that $Q\rightarrow 0$)
\beq
\label{eq:Lax operator U(N)}
\mathcal{R}(x)=\widetilde{\mathcal{R}}(x-i\frac{1}{\sqrt{N}}J_1)=\prod_{j=1}^N\left(x+\frac{1}{\sqrt{N}}\partial\fb+(\hs_j,\partial \bphi)\right)=\prod_{j=1}^N\left(x+\partial\varphi_j\right)\,,
\eeq
where we have used $\frac{1}{\sqrt{N}}\fb+(\hs_j, \bphi)=\varphi_j$. The currents $\mathcal{J}_s$ are given by expanding the Lax operator\footnote{The Toda action \eqref{eq:The Toda action} can be referred to as the $\SU(N)$ Toda CFT and the algebra $\bW_N$ as the $\SU(N)$ W-algebra. Adding the decoupled free boson $\fb$ brings us to the $\UU(N)$ Toda CFT and the currents \eqref{eq:defcalJ} generate the $\UU(N)$ W-algebra.} $\mathcal{R}(x)$. We get
\beq
\label{eq:defcalJ}
\mathcal{J}_s=\sum_{\ell=0}^s\binom{N-\ell}{N-s}W_{\ell}\left(\frac{-i}{\sqrt{N}} J_1\right)^{s-\ell}\,,
\eeq
with $W_0=1$ and $W_1=0$. In particular, one has $\mathcal{J}_1=-\tfrac{i}{\sqrt{N}}J_1$ for the normalized spin one current. One can of course derive the expressions for the currents $\mathcal{J}_s$ for general values of the shift $\kappa$, but we don't need them in what follows.
The relation between the currents $\mathcal{J}_s$ and the curve coefficients reads
\beq
\label{eq:identification calJ with cuco}
\vvac{\mathcal{J}_s(t)}_n=\cuco_s^{(n)}(t)\,.
\eeq
In order to have \eqref{eq:J1 and cuco1}, the primary fields have to also carry a $J_1$ charge $p$ as
\beq
\label{eq:complete vertex operators with fb}
\ver_{\astrosun}=e^{\form{\ba_{\astrosun}}{\bphi}}e^{p_{\astrosun}\fb}\,,\qquad \ver_{\bullet}=e^{\form{\ba_{\bullet}}{\bphi}}e^{p_{\bullet}\fb}\,.
\eeq
We can now compare $ \vvac{J_1(t)}_n$ with the SW curve coefficient $\cuco_1^{(n)}$ to fix the charges $p_{\astrosun}$ and $p_{\bullet}$. Let us consider the 4-point case. From \eqref{eq:curvecoefficientsSCQCDgeneralNandk} we get for $k=1$ and any $N$
\beq
\cuco_1^{(4)}(t)=\frac{q (M_L t+M_R (t-1))-M_L t^2+t u_1(q)}{(t-1) t (t-q)}\,.
\eeq
In order to make the coefficients of the highest order poles in $t$ independent of $q$, we need to set 
\beq
\label{eq:relation for u_1}
u_1(q)=\fa^{(1)}(1-q)\,,
\eeq
for $\fa^{(1)}$ defined in \eqref{eq:DefinitionCasimirs for the Coulomb}, which leads to $\cuco_1^{(4)}(t)=\frac{\fa^{(1)}-M_L}{t-1}+\frac{-\fa^{(1)}+M_R}{t-q}-\frac{M_R}{t}$. The $\UU(1)$ blocks needed for the computation of $\vvac{J_1}_4$ are found in appendix~\ref{app:U1blocks}. The comparison with \eqref{eq:vvacJ1for four points} tells us that  \eqref{eq:J1 and cuco1} is satisfied if we set the momenta of the vertex operators and intermediate state to
\beq
\label{eq:definitionpiforN1U1}
p_1=i\frac{ M_L}{\sqrt{N}}\,,\qquad p_2=-i\frac{M_L-\fa^{(1)}}{\sqrt{N}}\,,\qquad p_3=i\frac{M_R-\fa^{(1)}}{\sqrt{N}}\,,\qquad p_4=-i\frac{M_R}{\sqrt{N}}\,,\qquad p=-i\frac{\fa^{(1)}}{\sqrt{N}}\,.
\eeq
The above agrees with \eqref{eq:NekrasovConformalBlockU1} for $\epsilon_2=\epsilon_1^{-1}$ and in the limit $Q\rightarrow 0$. In the 3-point case, we have $p_1=i\tfrac{M_L}{\sqrt{N}}$, $p_2=-i\tfrac{M_L-M_R}{\sqrt{N}}$ and $p_3=-i\tfrac{M_R}{\sqrt{N}}$.

\paragraph{Comparisons with the curves. } We refer to appendix~\ref{app:Blocks} for the computations of the $\bW_2$ and $\bW_3$ blocks relevant for the comparison with the curve coefficients and to \cite{Mironov:2009dr} for an overview of the techniques needed for these computations. 

For the stress-energy tensor, we compute $\vvac{T(t)}_3$ in \eqref{eq:gvJ123 for T} and  $\vvac{T(t)}_4$ to quadratic order in $q$ in \eqref{eq:vvacTforN2}. Comparing them with $\tilde{\cuco}_2^{(n)}=\cuco^{(n)}_2-\tfrac{N-1}{2 N}(\cuco^{(n)}_1)^2$, with the $\cuco^{(n)}_s$ from \eqref{eq:curvecoefficientsSCQCDgeneralNandk},\eqref{eq:curvecoefficientstriniongeneralNandk}, leads to a perfect agreement if one sets the Coulomb branch parameter $u_2$ to be equal to\footnote{Observe that the transition from the SCQCD curve to the free trinion one makes us set $a_i=\mR{i}$, which puts $u_{\ell}(q=0)=(-1)^{\ell+1}\fc_R^{(\ell)}$, see \eqref{eq:weak coupling limit of ukl}, \eqref{eq:relation for u_1} and \eqref{eq: u2 related to CFT data}.} 
\begin{align}
\label{eq: u2 related to CFT data}
u_2(q)=&-\fa^{(2)}+\frac{q}{\tilde{\fa}^{(2)}}\Bigg[ -\frac{\fc_L^{(2)}\fc_R^{(2)}}{2}+ \frac{ (N-1) \fa^{(1)}(M_L \fc_R^{(2)}+\fc_L^{(2)} M_R)}{2 N}-\fa^{(2)}  \Big(\frac{N-1}{N}M_L M_R +\frac{\fc_L^{(2)}}{2}+\frac{\fc_R^{(2)}}{2}\Big)\nonumber\\&+\frac{(N-1) \fa^{(1)} \fa^{(2)}  (M_L+M_R)}{2 N}+\fa^{(2)} \Big(\frac{\fa^{(2)}}{2}-\frac{ N-1}{2 N}(\fa^{(1)})^2\Big)\Bigg]+\mathcal{O}(q^2)\,.
\end{align}

Similarly, $\vvac{W_3(t)}_3$ is to be found in \eqref{eq:3pointconformalblockwithWforW3general} and  $\vvac{W_3(t)}_4$ can be computed to linear order in $q$ with the tools provided in appendix~\ref{app: W3 blocks}. We compare them with $\tilde{\cuco}_3^{(n)}$, where
\beq
\tilde{\cuco}_3^{(n)}=\cuco_3^{(n)}-\frac{(N-2)  }{N}\cuco_1^{(n)}\cuco_2^{(n)}+\frac{(N-2) (N-1)}{3 N^2}(\cuco_1^{(n)})^3\,.
\eeq
The comparison works perfectly if we use the parameter identification of section~\ref{subsec:parameter identification} and if we express $u_3$ as a function of $q$, of the $\fa^{(s)}$ and of the mass parameters, just like we did for $u_2$ in \eqref{eq: u2 related to CFT data}. One can even perform the comparison for $\bW_4$, see \cite{Blumenhagen:1990jv} for the commutation relations, but the computations become very tedious and we omit them.

%----------------------------------------------------------------------------------------------
\section{The AGT correspondence for the $\calS_k$ theories.}
\label{sec: SkAGT correspondence}
%----------------------------------------------------------------------------------------------

Having reviewed in the last section some essential elements of the AGT correspondence, we can now apply them to the $\calS_k$ theories. \textit{The main principle guiding us is the observation that the class $\calS_k$ curves for $\SU(N)$ can be obtained from the $\calN=2$ $\calS$ curves for $\SU(Nk)$.}

In order to see that, we introduce a map that takes the $\SU(Nk)$ curve and sets the mass/Coulomb parameters to special values. Let us write this map as $\pro_{N,k}$ and define its action on  the $\SU(Nk)$ masses and Coulomb parameters as follows
\beq
\label{eq:defofpiNk}
\mL{j+N s}^{\SU(Nk)}\longmapsto \mL{j} \e^{\frac{2\pi i }{k}s}\,, \qquad \mR{j+N s}^{\SU(Nk)}\longmapsto \mR{j} \e^{\frac{2\pi i }{k}s}\,,\qquad 
a_{j+N s}^{\SU(Nk)}\longmapsto a_{j} \e^{\frac{2\pi i }{k}s}\,,
\eeq
where the indices run as $j=1,\ldots, N$, $s=0,\ldots, k-1$. The parameters on the right hand side of \eqref{eq:defofpiNk} are those of the class $\calS_k$ $\SU(N)$ theory.
Since $\prod_{s=0}^{k-1}\big(v-m \e^{\frac{2\pi i }{k}s}\big)=v^k-m^k$, 
it is clear from the curve equations \eqref{eq:curvewithgaugegroup} and \eqref{eq:curveTrinionGeneralOrbifold} that $\pro_{N,k}$ maps the $\calN=2$ $\SU(Nk)$ curve with $k=1$ to the $\calN=1$ $\calS_k$ $\SU(N)$ curve. Furthermore, it is clear that $\pi_{N,k}$ maps the sums of all the left/right masses to zero.
This generalizes to the following action on the Casimirs:
\beq
\label{eq: action of pi on the Casimirs}
\pro_{N,k}\left(\fc^{(k\ell ),\SU(Nk)}\right)=(-1)^{\ell(k+1)}\fc^{(\ell,k)}\,.
\eeq
and $\pro_{N,k}\left(\fc^{(s),\SU(Nk)}\right)=0$ if $s\neq  k\ell$. The above is proved in appendix~\ref{app:useful}, see equation \eqref{eq:sumsandCasimirs}. The action \eqref{eq: action of pi on the Casimirs} together with the expression for the $u_\ell$ as functions of the Casimirs (for example \eqref{eq:relation for u_1} and \eqref{eq: u2 related to CFT data}) implies that for $Q=0$ we have
\beq
u_s^{\SU(Nk)}\longmapsto \left\{\begin{array}{ll}u_{s}& \text{ if }s \text{ mod }k =0\\0& \text{ otherwise }\end{array}\right.\qquad \text{ with }  s=1,\ldots, Nk\,.
\eeq

\textbf{Our guiding principle can now be stated as follows:} since the map \eqref{eq:defofpiNk} sends the $\calN=2$ $\SU(Nk)$ curve to the $\calN=1$ $\SU(N)$ class $\calS_k$ curve, we can expect that $\pi_{N,k}$ would preserve the aspects of the AGT correspondence of section~\ref{sec: reviewAGT}, namely the identification of blocks and instanton partition functions as well as the correspondence between the curves and the ratios of the blocks with current insertions. 

In this section, we shall study the consequences of this principle. We begin with some $\bW_N$ representation theory and show in particular that the simple punctures are mapped by $\pi_{N,k}$ to non-unitary representations. Following that, we look at the structure of the corresponding 3 and 4-point blocks and study the Ward identities. Finally, we compute the corresponding $\vvac{W_s}_n$ in the limit $Q\rightarrow 0$ and recover the $\calS_k$ curves \eqref{eq:curvecoefficientsSCQCDgeneralNandk} and \eqref{eq:curvecoefficientstriniongeneralNandk}, thus providing a check of the proposal.

%----------------------------------------------------------------------------------------------
\subsection{The structure of the punctures}
\label{subsec:structuresimple}
%----------------------------------------------------------------------------------------------

Let us now study the consequences of the map \eqref{eq:defofpiNk} on the punctures. 

For $k=1$, the full punctures $\ver_{\astrosun}$ are generic $\bW_N$ representations with no special properties, while the simple ones $\ver_\bullet$ are representations with $\frac{(N-2)(N-1)}{2}$  null vectors, which allows us to compute the three and four point W-blocks. Both the simple and the full punctures are unitary representations of $\bW_N$. 

\paragraph{The simple punctures. }For $k>1$, all the charges of the simple punctures vanish, i.e.~$\bcharge_\bullet =\{0,\ldots, 0\}$. This follows from the fact that, see \eqref{eq:Simple punctures and masses}, the parameter $\varkappa$  determining $\ba_\bullet$ is given by the sum of all the left/right masses which are mapped by $\pi_{N,k}$ to zero. \textit{However, the} $\ver_\bullet$ \textit{are still different from the identity field} $\id$! The first and most important difference is that $L_{-1}\id= 0$ but $L_{-1}\ver_\bullet \neq 0$, because otherwise, the W-block would not depend on the insertion point of the simple puncture, which would prevent us from recovering the curve coefficients from $\vvac{W_s}_n$. Of course, the norm of the state $L_{-1}\ver_\bullet$ for $k>1$ must be zero, since $\bracket{L_{-1}\ver_\bullet}{L_{-1}\ver_\bullet}=2\cd_\bullet  \bracket{\ver_\bullet}{\ver_\bullet}$ and $\cd_\bullet$ is zero. Since we have non-zero states with zero norm, the CFT that we need to consider for the $\calS_k$ AGT correspondence \textbf{is non-unitary}.

\begin{figure}[htbp!]
             \begin{center}       
              \includegraphics[scale=0.25]{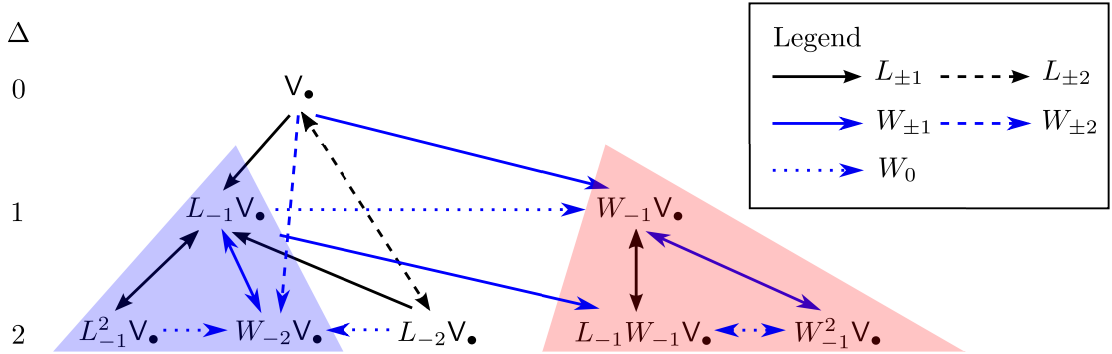}
              \caption{Structure of the first 3 levels of the simple puncture for $N =1$ and $k=3$ which implies $\cd=\charge=0$. For $c=2$, one quotients out the submodule (shaded in red) generated by $W_{-1}\ver_{\bullet}$. For $c\neq 2$, i.e.~for $Q\neq 0$, one should quotient out the submodule generated by the vector $(W_{-1}+\tfrac{Q}{2}L_{-1})\ver_{\bullet}$ instead. We remark that the singular vector $L_{-1}\ver_{\bullet}$ generates an indecomposable submodule, shaded in blue, whose elements all have zero norm. If we were to quotient out the zero norm states as well, then we would obtain the identity representation. The color and type of the of the arrows indicates which generators are acting, as depicted in the legend. }
              \label{Fig:simplepuncturestructure}
            \end{center}
\end{figure}

We can now look at the null states in the simple punctures. First, let us consider the case $Q=0$, which allows us to learn from the Seiberg-Witten curve. We see that the curve coefficients \eqref{eq:curvecoefficientsSCQCDgeneralNandk} have only simple poles at $t=1$ and $t=q$. For $k=1$, this is due to the presence of the $\UU(1)$ factors. In that case, we can shift $x\rightarrow x-\frac{1}{N}\cuco_1$ and then obtain curve coefficients $\tilde{\cuco}_\ell$ that have poles of order $\ell$ at $t=1$ and $t=q$ whose coefficients are related to the action of the modes $W_{\ell,-n}$ by \eqref{eq: action of W_s on a primary}. For $k>1$, we are not allowed to shift in $x$ anymore\footnote{By \eqref{eq:definitionpiforN1U1} the $\UU(1)$ charges are zero since $\pi (\fa^{(1)})=\pi (M_L)=\pi (M_R)=0$. Hence the $\UU(1)$ contribution is zero and is not responsible for the fact that the poles at the simple puncture are only first order.} and therefore, we have to conclude that  
\beq
\label{eq:simplepunctureconditionW}
W_{s,-n}\ver_\bullet=0\qquad \text{ for }n=0,1,\ldots, s -2\,,
\eeq
for all $s=2,\ldots, Nk$. This of course confirms that the charges $\bcharge$ of the simple puncture vanish and implies there are 
\beq
\sum_{s=2}^{Nk}(s-2)=\frac{(Nk-2)(Nk-1)}{2}
\eeq
null vectors. Hence, the number of null vectors for the simple punctures of the $\SU(N)$ $\calS_k$ theories is the same as for the $\calN=2$ $\SU(Nk)$ theories. Hence we conjecture that the null vectors are inherited from the $\calN=2$ theory, i.e.~obtained from it by mapping the parameters with $\pi_{N,k}$. Let us check this for the case $Nk=3$, where we write for simplicity $W_n\equiv W_{3,n}$ for the modes. For general $Q$ and $k$,  we can use \eqref{eq:defConformalDimensionsforToda}, \eqref{eq:chargeW3 for general Q} and \eqref{eq:Simple punctures and masses} to compute  for the simple  puncture $\cd_\bullet= \frac{1}{3} \varkappa  (3 Q-\varkappa )$, $\charge_\bullet= -\frac{1}{27} \varkappa (3 Q-\varkappa ) (3 Q-2 \varkappa ) $. Hence, the null vector is 
\beq
\label{eq:generalnullvectorN3}
\big(W_{-1}-\frac{3\charge_\bullet}{2\cd_\bullet}L_{-1}\big)\ver_\bullet=\big(W_{-1}+\frac{3Q-2\varkappa}{6}L_{-1}\big)\ver_\bullet=0\,.
\eeq
For $k>1$,  $\pi_{N,k}$ maps the parameter $\varkappa$ to zero and we have $\cd_\bullet=\charge_\bullet=0$. By \eqref{eq:generalnullvectorN3} the limit $\varkappa\rightarrow 0$ of the ratio $\frac{\charge_\bullet}{\cd_\bullet}$ is non-zero, leading to the null vector $\big(W_{-1}+\tfrac{Q}{2}L_{-1}\big)\ver_\bullet=0$. For $Q=0$, this gives (just like the curves do, see \eqref{eq:simplepunctureconditionW}) the condition $W_{-1}\ver_\bullet=0$, confirming the conjecture that the null vectors are inherited from the $\calN=2$ case.

Let us now show the structure of the simple puncture $\ver_\bullet$ in more detail, again taking the $\bW_3$ algebra case for simplicity. For further simplicity, we set $Q=0$ so that the null vector is $W_{-1}\ver_\bullet$.   The structure of the first three levels of the representation is depicted in figure~\ref{Fig:simplepuncturestructure}. It is important to remark that the structure shown in figure~\ref{Fig:simplepuncturestructure} holds only for $c=2$, i.e.~for $Q=0$. Otherwise, there are generators that act on the states like $W_{-1}^2\ver_\bullet$, that have to be set to zero, but don't give zero, meaning that the quotient is only well defined if $c=2$, i.e.~for $Q=0$. This is to be expected, since the null vector for $Q\neq 0$ is $\big(W_{-1}+\tfrac{Q}{2}L_{-1}\big)\ver_\bullet$. We remark that, unlike for generic $\bW_N$ Verma modules, the action of the $W_{s,0}$ modes with $s>2$ on the simple punctures will not be diagonalizable.

%-------------------------------------------------------------------------------------
\begin{table}[ht!]
\centering
\renewcommand{\arraystretch}{1.6}
\begin{tabular}{|c|c|c|}
\hline

& 
$\ver_\bullet$
& 
$\ver_{\astrosun}$
\\
\hline
$\cd$ for $k=2$ & 0   & $\frac{N (4N^2-1)}{12}Q^2-\sum_{i=1}^Nm_i^2$\\
\hline
$\cd$ for $k>2$ & 0   & $\frac{N k ((Nk)^2-1)}{24}Q^2$ \\
\hline Higher charges & 0& $\neq 0$\\
\hline
Null states for $Q=0$ & $W_{s,-n}\ver_\bullet=0$  for $n=0,1,\ldots, s -2$ and $s=2,\ldots, Nk$ & None\\
\hline
\end{tabular}
\renewcommand{\arraystretch}{1.0}
\caption{ This table contains an overview of the main properties of the punctures for the $\SU(N)$ $\calS_k$ theory for $k>1$.}
\label{tab:PunctureOverview}
\end{table}

\paragraph{The full punctures. }For $k>1$ and $Q=0$, the curve coefficients \eqref{eq:curvecoefficientsSCQCDgeneralNandk} imply that some of the charges of the full punctures $\ver_{\astrosun}$ become zero as well. Specifically, only the $\charge_{k\ell}$ with $\ell=1,\ldots, N$ are non-zero. For $k>2$, this implies that for $Q=0$ the conformal dimension of the full punctures vanishes, i.e.~$\cd_{\astrosun}=0$. However, we do not want the full punctures to become the identity field and hence, as for the simple punctures, we require that $L_{-1}\ver_{\astrosun}\neq 0$. Thus, they generically correspond to non-unitary representations as well, only without null-states. The main properties of the punctures are summarized for the reader's convenience in table~\ref{tab:PunctureOverview}.

\smallskip

We wish to finish this section with a remark. In the Toda theory, the primary fields, both those corresponding to the full punctures as well as those corresponding to the simple ones are obtained as $\ver=e^{\form{\ba}{\bphi}}$ for some appropriate $\ba$. In the CFTs that ought to be dual to the $\calN=1$ class $\calS_k$ theories, this is still true for the full punctures, \textit{but cannot be true for the simple ones} since for them the exponent is mapped by $\pi_{N,k}$ to zero and $e^0=\id$ is the identity field. It is unclear whether it is possible to write the simple punctures by using the Toda fields $\bphi$ at all.

%----------------------------------------------------------------------------------------------
\subsection{The 3-point blocks with one simple puncture}
\label{subsec: the 3 point blocks for the orbifold theory}
%----------------------------------------------------------------------------------------------

Let us now take the general considerations of the previous subsections and use them to compute the 3-point W-blocks. We perform the computations in the limit $Q\rightarrow 0$ that is needed for the comparison with the curves.
Let us denote by $\widehat{\ver}_{\bcharge}$ an arbitrary descendant of the primary $\ver_\bcharge$.
We compute using standard CFT techniques the recursion relations
 (each contour integral comes equipped with a factor of $\tfrac{1}{2\pi i}$ that we omit)
\beq
\begin{split}
\vac{\ver_1(\infty)\ver_2(1)(W_{s,-n}\widehat{\ver}_{\bcharge})(0)}=&\oint_0 \frac{dz}{z^{n-s+1}}\vac{\ver_1(\infty)\ver_2(1)W_{s}(z)\widehat{\ver}_{\bcharge}(0)}\\
=&-\sum_{k=-\infty}^{\infty}\oint_1\frac{dz}{z^{n-s+1}(z-1)^{k+s}}\vac{\ver_1(\infty)(W_{s,k}\ver_2)(1)\widehat{\ver}_{\bcharge}(0)}\\&+(-1)^s\sum_{k=-\infty}^{\infty}\oint_{\infty}dz\frac{z^{k-s}}{z^{n-s+1}}\vac{(W_{s,k}\ver_1)(\infty)\ver_2(1)\widehat{\ver}_{\bcharge}(0)}\,,
\end{split}
\eeq
where in the last line we have used (for a primary field) the relation $W_s(z^{-1})=(-z^{-2})^sW_s(z)$ and also the fact that the contour had to be oriented the other way. Computing the residues, we find for $n\geq 0$
\beq
\label{eq:recursionrelationWspart1}
\begin{split}
\vac{\ver_1(\infty)\ver_2(1)(W_{s,-n}\widehat{\ver}_{\bcharge})(0)}
=&(-1)^s\vac{(W_{s,n}\ver_1)(\infty)\ver_2(1)\widehat{\ver}_{\bcharge}(0)}-\vac{\ver_1(\infty)(W_{s,-s+1}\ver_2)(1)\widehat{\ver}_{\bcharge}(0)}\\&-\binom{-n+s-1}{s-1}\vac{\ver_1(\infty)(W_{s,0}\ver_2)(1)\widehat{\ver}_{\bcharge}(0)}\,,
\end{split}
\eeq
where we have used \eqref{eq:simplepunctureconditionW} following from the fact that $\ver_2$ is a simple puncture. At this point, there is a distinction between the case $k=1$ (i.e.~for $\calN=2$ gauge theories) in which $W_{s,-n}\ver_2$ can be expressed through the $L_{-m}\ver_2$ and the case $k>1$ (i.e.~$\calN=1$ gauge theories) in which $W_{s,-n}\ver_2=0$. We only consider the latter case here and write the recursion relations for $k=1$ and $N=2,3$ in appendix~\ref{app:Blocks}.  Plugging $n=0$ in \eqref{eq:recursionrelationWspart1}, we find the relation
\beq
\label{eq:special3ptidentityWs}
\begin{split}
\vac{\ver_1(\infty)(W_{s,-s+1}\ver_2)(1)\widehat{\ver}_{\bcharge}(0)}=&\big((-1)^s\charge_{s;1}-\charge_{s;2}\big)\vac{\ver_1(\infty)\ver_2(1)\widehat{\ver}_{\bcharge}(0)}\\&-\vac{\ver_1(\infty)\ver_2(1)(W_{s,0}\widehat{\ver}_{\bcharge})(0)}\,.
\end{split}
\eeq
In the above, we denote by $w_{s;i}$ the charge of $W_s$ when acting on the primary $\ver_i$.
Remark that the action of $W_{s,0}$ on descendant states does not need to be diagonal, unlike the action of $L_0$. Plugging \eqref{eq:special3ptidentityWs} into \eqref{eq:recursionrelationWspart1}, we obtain for $n>1$ the expression
\beq
\label{eq:recursionrelationWspart2}
\begin{split}
\vac{\ver_1(\infty)\ver_2(1)(W_{s,-n}\widehat{\ver}_{\bcharge})(0)}=&\vac{\ver_1(\infty)\ver_2(1)(W_{s,0}\widehat{\ver}_{\bcharge})(0)}\\&+\left[\left(1-\binom{-n+s-1}{s-1}\right)\charge_{s;2}-(-1)^s\charge_{s;1}\right]\vac{\ver_1(\infty)\ver_2(1)\widehat{\ver}_{\bcharge}(0)}\,.
\end{split}
\eeq
For the computation of 4-point blocks, we also need the recursion relations for the $\gvt$ vertices. Using the same tools, we can derive the following relation for $n>1$
\beq
\label{eq:bargamma relations for the orbifold theories}
\bracket{W_{s,-n}\widehat{\ver}}{\ver_3(1)\ver_4(0)}=\bracket{W_{s,0}\widehat{\ver}}{\ver_3(1)\ver_4(0)}+\left[\left(\binom{n+s-1}{s-1}-1\right)\charge_{s;3}-\charge_{s;4}\right]\bracket{\widehat{\ver}}{\ver_3(1)\ver_4(0)}\,.
\eeq
The action of $W_{s,0}$ on descendant fields needs to be computed using the appropriate $\bW$-algebra commutation relation, which then together with $\eqref{eq:bargamma relations for the orbifold theories}$ allows us to compute the $\gvt$ vertices.

Finally, for two full and one simple puncture (hence with $\charge_{s;2}=0$),  we can use \eqref{eq:recursionrelationWspart2} and obtain  the W-block with insertion of the current
\beq
\label{eq:3 point W blocks for the orbifold theory}
\begin{split}
\gvJ_{12\bcharge}(W_{s}(t);\emptyset)=&\sum_{n=-\infty}^0t^{-n-s}\frac{\vac{\ver_1(\infty)\ver_2(1)(W_{s;n}\ver_{\bcharge})(0)}}{\vac{\ver_1(\infty)\ver_2(1)\ver_{\bcharge}(0)}}\\=&t^{-s}\charge_{s;\bcharge}+\sum_{n=1}^{\infty}(\charge_{s;\bcharge}-(-1)^s\charge_{s;1})t^{n-s}=\frac{(-1)^s\charge_{s;1}  t-\charge_{s;\bcharge}}{t^s(t-1)}\,.
\end{split}
\eeq
We can immediately compare the above with the curve coefficients\footnote{Remember that for $k>1$, we cannot do a shift in $x$, $\cuco_1=0$ and hence there is no difference between $\cuco_s$ and $\tilde{\cuco}_s$.} $\cuco_s^{(3)}$ of \eqref{eq:curvecoefficientstriniongeneralNandk}. We see that for $s=k\ell$, we have to have $\charge_{s;1}=(-1)^{\ell(k+1)}\fc_L^{(\ell,k)}$ and $\charge_{s;\bcharge}=(-1)^{\ell}\fc_R^{(\ell,k)}$, while for $s\neq k\ell$ the charges have to vanish. This is in complete agreement with the parametrization \eqref{eq:TodachargesFullpunctures} (we can omit the tilde, since the sum of the left/right masses is zero for $k>1$) of the $\SU(Nk)$ theory with the action \eqref{eq: action of pi on the Casimirs} of the projection on the Casimirs. 

Hence, we conclude that the 3-point blocks of two full and one simple puncture with insertion of the $W_s$ current do reproduce the curve coefficients of the orbifold gauge theories if one uses the punctures of section~\ref{subsec:structuresimple}, i.e.~the punctures inherited from the $\SU(Nk)$ theory that have been acted upon by the projection $\pi_{N,k}$. 

\paragraph{Ward identities.} We can recover the formula \eqref{eq:3 point W blocks for the orbifold theory} also using Ward identities.  For a current $W_s$ of spin $sl$, we have the following Ward identities 
\beq
\label{eq:WardidentityforspinsCurrent}
\sum_{i=1}^n\left(\frac{W_{s,0;i}}{(t-z_i)^s}+\frac{W_{s,-1;i}}{(t-z_i)^{s-1}}+\cdots \frac{W_{s,-s+1;i}}{t-z_i}\right)\vac{\ver_1(z_1)\ldots \ver_n(z_n)}=0\,,
\eeq
where $W_{s,-m;i}$ is the mode $W_{s,m}$ acting on the $i^{\text{th}}$ field. Since we demand that $W_{s}(t)$ goes like $t^{-2s}$ at infinity, multiplying \eqref{eq:WardidentityforspinsCurrent} with $t^j$ with $j=0,\ldots, 2s-2$ and doing a contour integral around the insertion points of all the primary fields gives us $2s-1$ global Ward identities. We note that the $W_{s,0;i}$ act diagonally on the vertex operators, i.e. they just give the charges $\charge_{s;i}$. Let us summarize the counting of unknowns and constraints:
\begin{enumerate}
\item We have $2s-1$ independent Ward identities for an $n$-point function. The number is the same for any $n$.
\item For an $n$-point function, we have $n(s-1)$ unknowns that we need to determine in order to compute the ratio $\vac{W(t)\cdots }/\vac{\cdots}$ from \eqref{eq:WardidentityforspinsCurrent}. Each unknown corresponds to an insertion of a lowering operator $W_{s,-m}$ at the point $z_i$ in the correlation function, where $i\in\{1,2,\ldots, n\}$ and $m=1,\ldots, s-1$.
\item Since for the $n$-point function will have $n-2$ simple punctures, this gives through \eqref{eq:simplepunctureconditionW} exactly $(n-2)(s-2)$ conditions.
\end{enumerate}
In total, for an $n$-point function, we are left with
\beq
\label{eq:number of unknowns}
n (s-1)-(2 s-1)-(n-2) (s-2)=n-3
\eeq
unknowns. Thus, for $n=3$, we can compute the weighted correlation function with an insertion of the current just by using the Ward identities. We just need to insert the solutions for the unknowns in \eqref{eq:WardidentityforspinsCurrent}. Doing so, we obtain the same result as \eqref{eq:3 point W blocks for the orbifold theory}:
\beq
\frac{\vac{\ver_1(\infty)\ver_2(1)W_{s}(t)\ver_{\bcharge}(0)}}{\vac{\ver_1(\infty)\ver_2(1)\ver_{\bcharge}(0)}}=\gvJ_{12\bcharge}(W_{s}(t);\emptyset)\,.
\eeq
Thus, the comparison between the free trinion curve and the CFT data is trivial - it follows only from the assumptions for the full/simple punctures, their charges and the existence of the currents of appropriate spin. The appropriate form of the algebra becomes noticeable only at four points.

%----------------------------------------------------------------------------------------------
\subsection{Four point blocks and the instanton partition functions}
\label{sec:4ptblocks and instantons}
%----------------------------------------------------------------------------------------------

Having seen that the proposal we introduced at the beginning of the current section for the relationship between the CFT blocks and the orbifold $\calS_k$ curves works wonderfully for the case of three points, we now want to turn to the 4-point blocks. 

In the present section, we shall check our proposal by computing $\vvac{T(t)}_4\equiv \vvac{W_2(t)}_4$ for quadratic order in $q$ and $\vvac{W(t)}_4\equiv \vvac{W_3(t)}_4$ to linear order in $q$ for $k\geq 2$ and comparing to the curves.

%----------------------------------------------------------------------------------------------
\subsubsection{The four point blocks }
\label{subsubsec: the four point blocks orbifold}
%----------------------------------------------------------------------------------------------

In this section, we use \eqref{eq:general4ptvvacFormula} to compute $\vvac{W_s(t)}_4$. The relevant $\gv$ and $\gvt$ vertices are given either in the previous subsection~\ref{subsec: the 3 point blocks for the orbifold theory} or in appendix~\ref{app:Blocks}. 

\paragraph{The stress-energy tensor. } Let us consider first the case of the spin two current and compute $\vvac{T(t)}_4$ for the theories with $k\geq 2$. For $k=2$, we can simply take the general computation \eqref{eq:vvacTforN2} done in the appendix and set (use \eqref{eq:TodachargesFullpunctures}, $\ba_\bullet=0$ and \eqref{eq: action of pi on the Casimirs})
\beq
\cd_1=-\fc_L^{(1,2)}=-\sum_{i=1}^N\mL{i}^2\,,\quad \cd_2=\cd_3=0\,,\quad \cd_4=-\fc_R^{(1,2)}=-\sum_{i=1}^N\mR{i}^2\,,\quad \cd=-\fa^{(1,2)}=-\sum_{i=1}^Na_i^2\,.
\eeq
Plugging this in \eqref{eq:vvacTforN2}, we get the cumbersome expression for $\vvac{T(t)}_4$ up to quadratic order in $q$
\beq
\begin{split}
&\vvac{T(t)}_4=\frac{\fa^{(1,2)}-t \fc_L^{(1,2)}}{(t-1) t^2}-q \frac{ (\fa^{(1,2)}-\fc_R^{(1,2)}) ((t-2) \fa^{(1,2)}+t \fc_L^{(1,2)})}{2(t-1) t^3 \fa^{(1,2)}}\\&-q^2\frac{\fa^{(1,2)}-\fc_R^{(1,2)}}{ (t-1) t^4(2\fa^{(1,2)})^2\left(  c(1-2 \fa^{(1,2)})+2 \fa^{(1,2)} (8 \fa^{(1,2)}+5)\right)} \Bigg\{ c \Big[-(\fa^{(1,2)})^2 \left(4 t \fc_L^{(1,2)}+t^2 \fc_R^{(1,2)}-2 t+4\right)\\&+t \fa^{(1,2)} \fc_L^{(1,2)} (t \fc_L^{(1,2)}+2)-\left(t^2+4 t-8\right) (\fa^{(1,2)})^3+t^2 (\fc_L^{(1,2)})^2 \fc_R^{(1,2)}\Big]\\&+2 \fa^{(1,2)} \Big[(\fa^{(1,2)})^2 \left(t^2 (-2 \fc_L^{(1,2)}+\fc_R^{(1,2)}+2)+2 t (8 \fc_L^{(1,2)}+5)-20\right)\\&-t \fa^{(1,2)} \fc_L^{(1,2)} (t (\fc_L^{(1,2)}+6 \fc_R^{(1,2)}+2)-10)+\left(3 t^2+16 t-32\right) (\fa^{(1,2)})^3+5 t^2 (\fc_L^{(1,2)})^2 \fc_R^{(1,2)}\Big]\Bigg\}+\mathcal{O}\left(q^3\right)\,,
\end{split}
\eeq
where $c=2N-1$ is the central charge of the $\SU(2N)$ theory for $Q=0$. Comparing with $\cuco_2^{(4)}(t)$ (for $k=2$ and $N$ general) of \eqref{eq:curvecoefficientsSCQCDgeneralNandk}, we get a perfect agreement if the Coulomb modulus $u_2(q)$ takes the form
\beq
u_2(q)=\fa^{(1,2)}+\frac{q}{2}\left[\frac{\fc_L^{(1,2)} \fc_R^{(1,2)}}{\fa^{(1,2)}}+ (\fc_L^{(1,2)}+\fc_R^{(1,2)})-\fa^{(1,2)}\right]+\mathcal{O}(q^2)\,.
\eeq
Compare this result for $u_2(q)$ with the $k=1$ case of \eqref{eq: u2 related to CFT data}, while keeping the action \eqref{eq: action of pi on the Casimirs} in mind. In the above calculation, we computed $\vvac{T(t)}_4$ by doing the computation in the $\SU(2N)$ theory and then projecting using $\pi_{N,2}$. Alternatively, we can straightforwardly use the tools of the previous subsection~\ref{subsec: the 3 point blocks for the orbifold theory} and obtain the same result.

Since our proposal reproduces the curves, we are given hope that the blocks would give the $\calS_k$ instanton partition functions, even for $Q\neq 0$.  In particular, for $N=1$, the full algebra of the theory is $\bW_2$ and hence \eqref{eq:Virasoro 4 point block general} gives the full 4-point block. To first order in $q$, this  reads
\beq
\label{eq: four point block N1 and k2}
\block_{\cd}(\cd_1,\cd_2,\cd_3,\cd_4|q)=1-q\frac{2 (a^2 -M_L^2) (a^2 -M_R)}{4a^2-Q^2 }+\mathcal{O}(q^2)\,,
\eeq
since for $N=1$ we have $\cd=-a^2+\tfrac{Q^2}{4}$, $\cd_1=-M_L^2+\tfrac{Q^2}{4}$ and $\cd_4=-M_R^2+\tfrac{Q^2}{4}$, compare with table~\ref{tab:PunctureOverview}.

Computing $\vvac{T(t)}_4$ in the case $k>2$ is slightly trickier since for $Q=0$, the conformal dimension $\cd$ of the exchanged operator vanishes and one would need to divide by zero to compute the blocks. Hence, the correct approach is to perform the computation for $Q\neq 0$ such that $\cd=\frac{N k ((Nk)^2-1)}{24}Q^2$ (see table~\ref{tab:PunctureOverview}) and to then take the limit $Q\rightarrow 0$. This computation is well defined and it is straightforward to then check that $\lim_{Q\rightarrow 0}\vvac{T(t)}_4=0=\cuco_2^{(4)}(t)$, in agreement with \eqref{eq:curvecoefficientsSCQCDgeneralNandk}.

\paragraph{The spin three current. } The case of the $W_3$ current is straightforward too. For $k=3$ and $N$ general, the recursion relations of section~\ref{subsec: the 3 point blocks for the orbifold theory} give us after some straightforward computations
\beq
\begin{split}
&\gvJ_{12\bcharge}(W(t);\{\emptyset,\emptyset\})=\frac{-\charge_{1}  t-\charge_{\bcharge}}{(t-1)t^3}\,,\qquad 
\gvJ_{12\bcharge}(W(t);\{\{1\},\emptyset\})=\frac{(t-3) \charge_{\bcharge}-2 t \charge_1}{(t-1) t^4}\,, \\&\gvJ_{12\bcharge}(W(t);\{\emptyset,\{1\}\})=-\frac{(\charge_{\bcharge}+\charge_1) (t \charge_1+\charge_{\bcharge})}{(t-1) t^3}\,.
\end{split}
\eeq
Combined with $\gvt_{12\bcharge}(\{\{1\};\emptyset\})=0$, $\gvt_{12\bcharge}(\{\emptyset;\{1\}\})=\charge_{\bcharge}-\charge_4$ and \eqref{eq:W3shaplevel1} with $\cd_{\bcharge}=\cd_i=0$, we can calculate $\vvac{W(t)}_4$ to linear order in $q$. Since $\charge_1=\fc_L^{(1,3)}$, $\charge_{\bcharge}=-\fa^{(1,3)}$ and $\charge_4=-\fc_R^{(1,3)}$, we find
\beq
\begin{split}
\vvac{W(t)}_4&=\frac{1}{1+0\cdot q}\left[\frac{-\fc_L^{(1,3)}  t+\fa^{(1,3)}}{(t-1)t^3}+q\frac{1}{-3\fa^{(1,3)}}\frac{(3-t) \fa^{(1,3)}-2 t \fc_L^{(1,3)}}{(t-1) t^4}(-\fa^{(1,3)}+\fc_R^{(1,3)})\right]+\mathcal{O}(q^2)\,.
\end{split}
\eeq
The above agrees perfectly with the curve coefficient $\cuco_3^{(4)}(t)$ in  \eqref{eq:curvecoefficientsSCQCDgeneralNandk} for $k=3$ if we set the Coulomb modulus to the value
\beq
u_3(q)=\fa^{(1,3)}+\frac{q}{3}\left[\frac{2 \fc_L^{(1,3)} \fc_R^{(1,3)}}{\fa^{(1,3)}}+ (\fc_L^{(1,3)}+\fc_R^{(1,3)})-\fa^{(1,3)}\right]+\mathcal{O}(q^2)\,.
\eeq
Hence, our proposal agrees with the first non-trivial $\calS_3$ curve coefficient.

We can also compute (for $N=1$ and $k=3$) the 4-point block $\block$ for general $Q$.  The non-trivial $\bW_3$ charges are $\bcharge_1=\{Q^2,M_L^3\}$, $\bcharge_4=\{Q^2,-M_R^3\}$ and $\bcharge=\{Q^2,-a^3\}$ for the intermediate state. 
From \eqref{eq:W3shaplevel1}, we find after putting $c=2(1+12Q^2)$ for the first level Shapovalov form
\beq
\label{eq:shapW3leve1orbifoldN1k3}
\shap^{(1)}_{\bcharge}=\left(
\begin{array}{cc}
 2 Q^2 & -3 a^3 \\
 -3 a^3 & -\frac{Q^4}{6} \\
\end{array}
\right)\,.
\eeq
Since $\gv_{12\bcharge}(\{\{1\};\emptyset\})=\cd+\cd_2-\cd_1=Q^2-Q^2=0$ and (see \eqref{eq: W3 recursionrelationWs}) $\gv_{12\bcharge}(\{\emptyset;\{1\}\})=\charge_1+\charge_{\bcharge}=-a^3+M_L^3$. Similarly, see \eqref{eq:bargamma relations for the orbifold theories}, $\gvt_{12\bcharge}(\{\{1\};\emptyset\})=0$ and $\gvt_{12\bcharge}(\{\emptyset;\{1\}\})=-a^3+M_R^3$.  Hence, inverting \eqref{eq:shapW3leve1orbifoldN1k3}, we find that the $\bW_3$ block up to level 1 is 
\beq
\label{eq:blockforW3forN1k3}
\begin{split}
\block_{\bcharge}(\bcharge_1,\bcharge_2,\bcharge_3,\bcharge_4|q)&=1+q\left(-\frac{6 Q^2}{27 a^6+Q^6}\right)(-a^3+M_L^3)(-a^3+M_R^3)+\mathcal{O}(q^2)\\
&=1-q\frac{6 Q^2(a^3-M_L^3)(a^3-M_R^3)}{27 a^6+Q^6}+\mathcal{O}(q^2)\,.
\end{split}
\eeq

In addition to the computations for $\vvac{W_2}$ and $\vvac{W_3}$ that we have shown here, we have performed additional checks - for $\vvac{W_4}$ and for higher orders in $q$.

%----------------------------------------------------------------------------------------------
\subsubsection{The instanton partition function of the orbifold theories}
%----------------------------------------------------------------------------------------------

Having checked in the previous subsection that our proposal reproduces the curves, we now want to investigate the instanton partition functions. Since the AGT correspondence holds in $\calN=2$ case, it is trivial that the correspondence between the four-point blocks $\block$ of section~\ref{subsubsec: the four point blocks orbifold} will agree with the Nekrasov partition functions projected with $\pi_{N,k}$. Still, it is worth looking at the way the projection $\pi_{N,k}$ acts to see what we can learn from it about the class $\calS_k$ theories.

The image of the Nekrasov instanton partition function $\calZ_{\text{inst}}^{(Nk,1)}$ of the $\SU(Nk)$ $\calN=2$ SCQCD \eqref{eq:general Z inst} under the map $\pi_{N,k}$ can be easily obtained. We can use $\prod_{r=0}^{k-1}\big(a-m\e^{\tfrac{2\pi i }{k}r}\big)=a^k-m^k$ to write
\beq
\label{eq:general Z inst k and N}
\begin{split}
\calZ_{\text{inst}}^{(N,k)}=&\pi_{N,k}(\calZ_{\text{inst}}^{(Nk,1)})\define\sum_{\bY=\{Y_1,\ldots, Y_{Nk}\}}q^{|\bY|}\tilde{z}_{\text{inst}}^{(N,k)}(\bY)\\
=&\sum_{\bY=\{Y_1,\ldots, Y_{Nk}\}}q^{|\bY|}\prod_{u=1}^N\prod_{i=1}^N\prod_{r=0}^{k-1}\prod_{(\mu,\nu)\in Y_{i+N r}}\left[\left(\epsilon-a_i\e^{\tfrac{2\pi i }{k}r}-\epsilon_1\mu-\epsilon_2\nu\right)^k-\mL{u}^k\right]\\&\times \prod_{u=1}^N\prod_{i=1}^N\prod_{r=0}^{k-1}\prod_{(\mu,\nu)\in Y_{i+N r}}\left[\left(a_i\e^{\tfrac{2\pi i }{k}r}+\epsilon_1\mu+\epsilon_2\nu\right)^k-\mR{u}^k\right]\\&\times \Bigg\{\prod_{i,j=1}^N\prod_{r,s=0}^{k-1}\prod_{(\mu,\nu)\in Y_{i+N r}}\left[a_i\e^{\tfrac{2\pi i}{k}r}-a_j\e^{\tfrac{2\pi i}{k}s}-\epsilon_1L_{Y_{j+Ns}}(\mu,\nu)+\epsilon_2\left(A_{Y_{i+Nr}}(\mu,\nu)+1\right)\right]\\&\times \prod_{(\mu',\nu')\in Y_{j+N s}}\left[\epsilon+a_i\e^{\tfrac{2\pi i}{k}r}-a_j\e^{\tfrac{2\pi i}{k}s}+\epsilon_1L_{Y_{i+Nr}}(\mu',\nu')-\epsilon_2\left(A_{Y_{j+Ns}}(\mu',\nu')+1\right)\right]\Bigg\}^{-1}\,.
\end{split}
\eeq
The resulting sum is still full of phases which lead to many cancellations when the sums over the partitions are performed. It is useful to split the sum over the partitions $\bY$ into orbits of the orbifold group $\mathbb{Z}_N$, where the action of that group on $\bY$ is defined via the elementary cyclic shift
\begin{multline}
\{Y_1,\ldots, Y_N, Y_{N+1},\ldots, Y_{2N},\ldots, Y_{(k-1)N+1},\ldots, Y_{kN}\}\longmapsto\\\longmapsto \{Y_{(k-1)N+1},\ldots, Y_{kN}, Y_1,\ldots, Y_N, \ldots, Y_{(k-2)N+1},\ldots, Y_{(k-1)N}\}\,.
\end{multline}
Thus, we can rewrite the instanton partition function with the summands expressed as sums over the cyclic permutations:
\beq
\label{eq:instanton partition function orbifold}
\begin{split}
\calZ_{\text{inst}}^{(N,k)}&=\sum_{\bY=\{Y_1,\ldots, Y_{Nk}\}}q^{|\bY|}\tilde{z}_{\text{inst}}^{(N,k)}(\bY)=\sum_{[\bY]\in \{Y_1,\ldots, Y_{Nk}\}/\mathbb{Z}_k}q^{|\bY|}\underbrace{\sum_{\sigma\in \mathbb{Z}_k}\tilde{z}_{\text{inst}}^{(N,k)}(\sigma\cdot \bY)}_{\define z_{\text{inst}}^{(N,k)}([\bY])}\,.
\end{split}
\eeq
It seems quite non-trivial to obtain closed analytic expressions for the $z_{\text{inst}}^{(N,k)}([\bY])$ for general $N$, $k$ and equivalence class $[\bY]$. For the simplest case of $N=1$ and $k$ general, one finds for the first non-trivial equivalence class $[\{\{1\},\emptyset,\ldots, \emptyset\}]$ the result
\beq
\label{eq: first zinst}
z_{\text{inst}}^{(1,k)}\big([\{\{1\},\emptyset,\ldots, \emptyset\}]\big)=-\frac{\epsilon \left(a^k-M_L^k\right)}{\epsilon_1\epsilon_2 k a^{k-1}} \sum _{s=0}^{k-1} \e^{\frac{2 \pi  i}{k}s} \frac{\left(\epsilon+a \e^{\frac{2 \pi  i }{k}s}\right)^k-M_R^k}{\left(\epsilon+a \e^{\frac{2 \pi  i }{k}s}\right)^k-a^k}\,.
\eeq
The first few cases of $z_{\text{inst}}^{(1,k)}\equiv z_{\text{inst}}^{(1,k)}\big([\{\{1\},\emptyset,\ldots, \emptyset\}]\big)$ with $k>1$ can be simplified to 
\begin{align}
&z_{\text{inst}}^{(1,2)}=-\frac{2 \left(a^2-M_L^2\right) \left(a^2-M_R^2\right)}{\epsilon_1\epsilon_2(4 a^2-\epsilon^2)}\,,&
&z_{\text{inst}}^{(1,3)}=-\frac{6 \epsilon^2 \left(a^3-M_L^3\right) \left(a^3-M_R^3\right)}{\epsilon_1\epsilon_2(27 a^6+\epsilon^6)}\,,&\nonumber\\
&z_{\text{inst}}^{(1,4)}=\frac{20 \epsilon^2 \left(a^4-M_L^4\right) \left(a^4-M_R^4\right)}{\epsilon_1\epsilon_2(-64 a^8-12 a^4 \epsilon^4+\epsilon^8)}\,,&
&z_{\text{inst}}^{(1,5)}=-\frac{10 \epsilon^2  \left(125 a^{10}+7 \epsilon^{10}\right)\left(a^5-M_L^5\right) \left(a^5-M_R^5\right)}{\epsilon_1\epsilon_2(3125 a^{20}+625 a^{10} \epsilon^{10}+\epsilon^{20})}\,.&
\end{align}
The above clearly agrees with \eqref{eq: four point block N1 and k2} and \eqref{eq:blockforW3forN1k3}.  We have checked for higher $k$ that for $k>1$ equation \eqref{eq: first zinst} is equal to $\frac{1}{\epsilon_1\epsilon_2 }\tfrac{P_k(\epsilon,a)}{P'_k(\epsilon,a)} \left(a^k-M_L^k\right) \left(a^k-M_R^k\right)$, where $P_k$ and $P'_k$ are homogeneous polynomials in $\epsilon$ and $a$ with $\text{deg}P'_k-\text{deg}P_k=2(k-1)$. 

In conclusion, we see that the Nekrasov partition function \eqref{eq:instanton partition function orbifold} does indeed reproduce the CFT blocks with non-unitary fields. It still remains to determine closed formulas for the summands $z_{\text{inst}}^{(N,k)}([\bY])$ that do not depend on the phases introduced by $\pi_{N,k}$.

%----------------------------------------------------------------------------------------------
\section{Conclusion and Outlook}
\label{app:conclusion and outlook}
%----------------------------------------------------------------------------------------------

In this article, we showed that the Seiberg-Witten curves of the $\SU(N)$ class $\calS_k$ gauge theories  derived in  \cite{Coman:2015bqq}
can be obtained from the weighted current correlation functions $\vvac{W_s(t)}$ of the $\bW_{Nk}$ algebra once the mass parameters of the $\SU(Nk)$ theory have been properly identified under the $\mathbb{Z}_k$ orbifold condition. To do this, we first found the quantum numbers of the vertex operators $\ver_{\astrosun}$ and $\ver_{\bullet}$ of the full and the simple punctures respectively, and observed that in general the punctures correspond to non-unitary representations of $\bW_{Nk}$. We then argued that the null vectors of the simple punctures are inherited from the $\SU(Nk)$ and performed several checks of our proposal by computing $\vvac{W_s(t)}_n$ for $s=2,3$ and both $n=3$ and $n=4$ points and comparing with the meromorphic differentials of the Seiberg-Witten curve.  We furthermore conjectured that the $\SU(Nk)$ Nekrasov instanton partition functions with the orbifold values of the masses and the Coulomb branch parameters \eqref{eq:general Z inst k and N} give the instanton contributions of the $\SU(N)$ class $\calS_k$ gauge theories. Moreover, it is natural to further conjecture that the algebra, the blocks and the instanton partition functions of any theory in class $\calS_\Gamma$ is also obtained in this way, with the masses and the Coulomb branch parameters identified under the $\Gamma \in $ADE orbifold condition.

It seems natural to think that the full extend of the AGT correspondence applies to the  class $\calS_\Gamma$ gauge theories. A necessary first step involves the computation of the full 3-point functions of two full and one simple puncture, which can then be used through a block decomposition \`a la \eqref{eq:blockdecomposition of the 4 point function} to compute the full 4-point CFT correlation function. This correlation function should correspond to the $S^4$ partition function of the  $\SU(N)$ class $\calS_k$ theories. For the 3-point functions of two full punctures and one simple one, the appropriate 4D theory is a free one, namely the orbifold of the free trinion:
\beq
\calZ^{S^4}_{\text{ free trinion}} =  \vac{\ver_{\astrosun}(\infty)\ver_\bullet(1) \ver_{\astrosun}(0)}  \,.
\eeq
Since we are dealing with a free theory, the $S^4$ partition function can be straightforwardly computed by counting the eigenvalues of Dirac and Laplace operators. This is work in progress \cite{Carstensen}. Once these 3-point correlation functions have been computed, one also needs to check that the 4-point function satisfies the CFT crossing relations.

For $\calN=2$ gauge theories in 4D, the $S^4$ partition function is not scheme independent \cite{Gerchkovitz:2014gta} and the scheme dependence is understood as transformations of the K\"ahler potential of the conformal manifold. For theories with only $\calN=1$ supersymmetry, the ability to control this ambiguity is lost\footnote{Despite these ambiguities, the partition functions still contain well defined physical information. For example, certain derivatives of the free energy are scheme independent \cite{Bobev:2013cja, Bobev:2016nua}.} \cite{Gerchkovitz:2014gta}. However, for theories in the class $\calS_\Gamma$ at the orbifold point we expect that to not be the case. 
Our expectations stem from the AdS/CFT correspondence, the inheritance arguments of \cite{Bershadsky:1998mb,Bershadsky:1998cb}
and our large experience from the study of  $\calN=2$ orbifold daughters of $\calN=4$ SYM \cite{Gadde:2009dj, Gadde:2010zi,Liendo:2011xb,Pomoni:2011jj,Pomoni:2013poa,Mitev:2014yba,Mitev:2015oty}. 
When all the coupling constants are equal to each other (i.e.~at the orbifold point), certain observables in the untwisted sector are equal to the $\calN=4$ ones. Since, the theories in class $\mathcal{S}_k$ are also orbifolds of $\calN=4$ SYM, the inheritance arguments apply to them. In addition, they are by definition orbifolds of the $\calN=2$ class $\mathcal{S}$ theories and we are studying the case with all the coupling constants equal. Hence, we expect certain observables to be equal to the corresponding $\calN=2$ ones as well and conjecture that the partition function on $S^4$ is well defined.

Our results so far suggest, with a bit of optimism,  that  for any supersymmetric theory with a Lagrangian description and an {\it abelian Coulomb phase}, we should be able to {\it guess the dual 2D CFT}, just by knowing: 1) {\it the Seiberg-Witten curve} from which one extracts the symmetry algebra, the representations and then the instanton partition functions and 2) {\it the free trinion partition function}. Once these two are known, it should be possible to compute the complete 3-point functions and to check that the 4-point function satisfies the crossing equations. 

Beyond this point, there are still many questions left open. Some of them concern exploring the nature of the CFTs dual to the $\calN=1$ class $\calS_k$ theories and, in particular, their marginal deformations. In a work in progress \cite{Tom}, the SW curves away from the orbifold point are investigated. It would be very important to find the 2D CFT operation that is dual to adding a marginal deformation to the orbifold point Lagrangian. 

In addition, it would be instructive to try to repeat  for the $\calN=1$ theories of class $\calS_\Gamma$ the strategy of \cite{Cordova:2016cmu}, who starting from the (2,0) theory in 6D where able to obtain a direct derivation of the AGT correspondence. In particular, it would be interesting to see what is the orbifolded version of the intermediate complex Chern-Simons theory in this approach. 

Since we conjectured in section~\ref{sec: SkAGT correspondence} that the instanton partition functions of the class $\calS_k$ theories are obtained from the $\calN=2$ ones after specializing the parameters, it would be very important to compute these instanton contributions from first principles following \cite{Dorey:2002ik}. Alternatively, one could try to adapt Nekrasov localization techniques \cite{Nekrasov:2002qd, Nekrasov:2003rj} and especially their most modern incarnation \cite{Nekrasov:2016ydq}.

In this article, we studied the effect of performing a $\mathbb{Z}_k$ orbifold on the transverse directions of the M5 branes that breaks the supersymmetry of the gauge theory down to $\calN=1$. This should be distinguished from quotienting out a $\mathbb{Z}_r$ on space time directions and considering the $\calN=2$ theory on $\mathbb{R}^4/\mathbb{Z}_r$. In the latter case, the dual CFT is a coset model (parafermionic Toda CFTs) and the correspondence has been studied in  \cite{Belavin:2011pp,Nishioka:2011jk,Bonelli:2011jx,Bonelli:2011kv,Wyllard:2011mn, Alfimov:2011ju, Belavin:2011sw} 
among others. It would be interesting to do both quotients, i.e.~to investigate the AGT correspondence for the class $\calS_k$ theories on $\mathbb{R}^4/\mathbb{Z}_r$.

One is also interested in more general correlation and partition functions. For the $\calN=2$ theories, the free trinion partition function only gives the 3-point correlation functions (i.e.~the 3-point structure constants) with one simple puncture, which is a semi-degenerate field. In order to compute the correlation functions of three generic fields, dual to the partition function of the full trinion $T_N$, we used the refined topological string vertex in \cite{Bao:2013pwa,Mitev:2014isa, Isachenkov:2014eya}. It would be important to develop the refined topological vertex for D-brane configurations subjected to the orbifold identification \eqref{OrbifoldAction}, for it would give us a path towards the 3-point correlation functions of arbitrary primary fields.

Another potential direction of investigation concerns supersymmetric line and surface operators/defects. It would be important to classify them for  the  class $\calS_\Gamma$ gauge theories and to understand precisely how they are realized in the 2D CFT side, following closely the work of  \cite{Alday:2009fs} for the $\calN=2$ case. See also the more recent reviews \cite{Okuda:2014fja,Gukov:2014gja} and references therein.  It seems very possible  that the results of the present paper will immediately apply. Furthermore, it would be important to make contact with the recent works of \cite{Ito:2016fpl, Maruyoshi:2016caf, Yagi:2017hmj} based on the superconformal index. 

Lastly, we would like to state that the existence of a dual CFT whose correlation functions reproduce the partition functions gives one hope that a generalization of Pestun's localization to some $\calN=1$ theories on $\mathbb{S}^4$ or the ellipsoid should be possible. This is currently being researched \cite{Costis}.

%----------------------------------------------------------------------------------------------
\section*{Acknowledgments}
%----------------------------------------------------------------------------------------------

We have greatly profited from discussions with Jan Peter Carstensen, Ioana Coman, Yannick Linke, Volker Schomerus and J\"org Teschner. We are particularly grateful to Futoshi Yagi for critically reading the manuscript and for suggesting several improvements. The work of EP is funded
by DFG via the Emmy Noether Programme ``Exact results in Gauge theories''.

%----------------------------------------------------------------------------------------------
%Appendix
%----------------------------------------------------------------------------------------------
\appendix

%!TEX root = ../WBlocks.tex
%%%%%%%%%%%%%%%%%%%%%%%%%%%%%%%%%%%%%%%%%%%%%

%----------------------------------------------------------------------------------------------
\section{Summation identities}
\label{app:useful}
%----------------------------------------------------------------------------------------------

The Casimirs are defined as (We write $\fc^{(s)}\equiv \fc^{(s,1)}$)
\beq
\label{eq:DefinitionCasimirs}
\fc^{(s,k)}=\sum_{i_1<\cdots < i_s=1}^Nm_{i_1}^k\cdots  m_{i_s}^k\,,\qquad \fc^{(0,k)}=1\,.
\eeq
For $k=1$, they obey the important identity allowing to express the Casimirs of $\SU(N)$ in terms of the $\UU(N)$ ones:
\beq
\label{eq:relationSUNandUNCasimirs}
\sum_{j=0}^i\frac{(-1)^{i-j}}{N^{i-j}}\binom{N-j}{N-i}\fc^{(j)}(\fc^{(1)})^{i-j}={\fc^{(i)}}_{\big| m_a\rightarrow \tilde{m}_a}\define \tilde{\fc}^{(i)}\,.
\eeq
We remind that $\tilde{m}_a=m_a-\frac{M}{N}$ with $M=\fc^{(1)}=\sum_{a=1}^Nm_a$. It is clear from the definition that $\tilde{\fc}^{(1)}=0$.

We have ($\car_{ij}$ is the $\SU(N)$ Cartan matrix) the following formulas for contractions involving the Cartan matrix and the fundamental weights
\beq
\label{eq:summationFormula1}
\begin{split}
\sum_{i_1,i_2=1}^{N-1}(\omega_{i_1},\omega_{i_2})\car_{i_1,i_2}&=N-1\,,\qquad 
\sum_{i_1,i_2,i_3,i_4=1}^{N-1}(\omega_{i_1},\omega_{i_2})(\omega_{i_3},\omega_{i_4})\car_{i_1,i_3}\car_{i_2,i_4}=N-1\,.
\end{split}
\eeq
The second identity follows from the first one if we also apply the first of the  formulas
\beq
\label{eq:summationFormula2}
\sum_{i,j=1}^{N-1}(\balpha,\omega_i)(\bbeta,\omega_j)\car_{i,j}=(\balpha,\bbeta)\,,
\qquad \sum_{i<j=1}^N(\balpha,\hs_i)(\bbeta,\hs_j)=-\frac{1}{2}(\balpha,\bbeta)\,.
\eeq
Finally, we have the following summation identity
\beq
\label{eq:sumsandCasimirs}
\sum_{(i_1,n_1)<\cdots <(i_\ell,n_\ell)}m_{i_1}e^{\frac{2\pi i n_1}{k}}\cdots m_{i_\ell}e^{\frac{2\pi i n_\ell}{k}}=(-1)^{(k+1)s}\sum_{i_1<\cdots<i_s=1}^Nm_{i_1}^k\cdots  m_{i_s}^k\,,
\eeq
if $\ell=ks$ with $s=0,1,\ldots$ and is zero otherwise. In the sum, the indices $i_j$ run over $1,\ldots, N$ and $n_j$ over $1,\ldots, k$ with the inequality $(i,h)<(i',n')$ iff $i<i'$ or $i=i'$ and $n<n'$. Equation \eqref{eq:sumsandCasimirs} is proven by expanding the left hand side of the identity $\prod_{n=1}^k\left(x-e^{\frac{2\pi i n}{k}}\right)=x^k-1$ in powers of $x$, which leads to the formula
\beq
\sum_{n_1<n_2<\cdots<n_l}^ke^{\frac{2\pi i }{k}(n_1+\cdots+n_l)}=\left\{\begin{array}{ll} 0 & \text{ if } l\neq k\\
(-1)^{k+1}& \text{ if }l=k\end{array}\right.\,.
\eeq
It hence follows that in the sum of \eqref{eq:sumsandCasimirs} only those terms remain for which the $i_j$'s clump into bunches of size $k$ for which the sum over the $n$'s gives  a factor of $(-1)^{k+1}$. This completes the proof of \eqref{eq:sumsandCasimirs}.

%----------------------------------------------------------------------------------------------
\section{Shapovalov forms}
\label{app:shapovalov}
%----------------------------------------------------------------------------------------------

%----------------------------------------------------------------------------------------------
\paragraph{The Virasoro case.} The Shapovalov form for the first 3 levels reads $\shap_\cd^{(0)}=\left(
1\right)$, $\shap_\cd^{(1)}=\left(2 \cd \right)$
as well as
\beq
\label{eq:shapLiouvillelevel23}
\begin{split}
\shap_\cd^{(2)}&=\left(
\begin{array}{cc}
 \frac{1}{2} (c+8 \cd ) & 6 \cd  \\
 6 \cd  & 4 \cd  (2 \cd +1) \\
\end{array}
\right)\,,\\
\shap_\cd^{(3)}&=\left(
\begin{array}{ccc}
 2 (c+3 \cd ) & 2 (c+8 \cd ) & 24 \cd  \\
 2 (c+8 \cd ) & c (\cd +2)+2 \cd  (4 \cd +17) & 36 \cd  (\cd +1) \\
 24 \cd  & 36 \cd  (\cd +1) & 24 \cd  \left(2 \cd ^2+3 \cd +1\right) \\
\end{array}
\right)\,.
\end{split}
\eeq
The last matrix is wrt.~to the basis $\{3\},\{1,2\},\{1,1,1\}$, where $\{1,2\}$ stands for $L_{-1}L_{-2}\ver_{\cd}$.
We remind that the generators in the algebra are ordered as $L_{-n_1}^{m_1}\cdots L_{-n_s}^{m_s}\ver_{\cd}$ with $n_i<n_{i+1}$. 

%----------------------------------------------------------------------------------------------
\paragraph{The $\bW_3$ case.} For the $\bW_3$, using the commutation relations of appendix~\ref{app:W3 algebra}, the first non-trivial Shapovalov form reads 
\beq
\label{eq:W3shaplevel1}
\begin{split}
\shap_{\cd,w}^{(1)}&=\left(
\begin{array}{cc}
 2 \cd & 3 w \\
 3 w & \frac{1}{48} (c-32 \cd -2) \cd  \\
\end{array}
\right)\,,
\end{split}
\eeq
in the basis $\{\{1\};\emptyset\}\equiv L_{-1}\ver_{\cd,w}$ and $\{\emptyset;\{1\}\}\equiv W_{-1}\ver_{\cd,w}$. Similarly, 
in the basis $\{\{2\};\emptyset\}$, $\{\{1,1\};\emptyset\}$, $\{\emptyset;\{2\}\}$ $\{\emptyset;\{1,1\}\}$, $\{\{1\};\{1\}\}$ we find at level 2
\beq
\label{eq:W3shaplevel2}
\begin{split}
\shap_{\cd,w}^{(2)}=&\left(
\begin{array}{ccc}
 \frac{1}{2} (c+8 \cd ) & 6 \cd  & 6 w  \\
 6 \cd  & 4 \cd  (2 \cd +1) & 12 w \\
 6 w & 12 w & -\frac{1}{6} \cd  (c+8 \cd +6) \\
 \frac{5}{48} (c-32 \cd -2) \cd  & 18 w^2+\frac{1}{8} \cd  (c-32 \cd -2) & -\frac{1}{8} w (c+48 \cd +14)  \\
 9 w & 6 (2 \cd  w+w) & \frac{1}{12} (c-32 \cd -2) \cd   \\
\end{array}
\right.
\\&
\left.
\begin{array}{cc}
 \frac{5}{48} (c-32 \cd -2) \cd  & 9 w \\
18 w^2+\frac{1}{8} \cd  (c-32 \cd -2) & 6 (2 \cd  w+w) \\
-\frac{1}{8} w (c+48 \cd +14) & \frac{1}{12} (c-32 \cd -2) \cd  \\
 \frac{(c-32 \cd -2) \cd  \left(-64 \cd ^2+2 (c-34) \cd +c-34\right)-27648 w^2}{2304} & \frac{1}{16} w (c-32 \cd -2) (2 \cd +3) \\
\frac{1}{16} w (c-32 \cd -2) (2 \cd +3) & \frac{1}{24} \left(216 w^2+\cd  (\cd +1) (c-32 \cd -2)\right) \\
\end{array}
\right)\,.
\end{split}
\eeq

%----------------------------------------------------------------------------------------------
\section{The $\bW_3$ algebra}
\label{app:W3 algebra}
%----------------------------------------------------------------------------------------------

We have $c=2(1+12Q^2)$ and introduce the parameter $\beta=\frac{16}{22+5c}=\frac{2}{4+15 Q^2}$\,. The commutation relations of the modes are
\beq
\begin{split}
\com{L_m}{L_n}\,=\,&\frac{c}{12}m(m^2-1)\delta_{m+n,0}+(m-n)L_{m+n}\\
\com{L_m}{W_n}\,=\,&(2m-n)W_{m+n}\\
\com{W_m}{W_n}\,=\,&-\frac{1}{3\beta}\frac{c}{3\cdot 5!}m(m^2-1)(m^2-4)\delta_{m+n,0}\\&\,-\frac{(m-n)}{3\beta}\left(\frac{(m+n+3)(m+n+2)}{15}-\frac{(m+2)(n+2)}{6}\right)L_{m+n}-\frac{(m-n)}{3}\Lambda_{m+n}\,,
\end{split}
\eeq
where the spin four field $\Lambda(z)=(TT)(z)-\tfrac{3}{10}\partial^2T$ has the mode expansion
\beq
\Lambda_m=\sum_{p=-\infty}^{-2}L_{p}L_{m-p}+\sum_{p=-1}^{\infty}L_{m-p}L_p-\frac{3}{10}(m+2)(m+3)L_m\,.
\eeq
Compared to the commutation relations given in \cite{Belavin:2016qaa}, we have rescaled $W\rightarrow iW$. The conformal dimension and $w$ charge are given by in terms of $\SU(3)$ weights through
\beq
\label{eq:defchargew3}
\cd(\ba)=\frac{\form{2\fQ-\ba}{\ba}}{2}\,,\qquad w(\ba)=-\form{\ba-\fQ}{\hs_1}\form{\ba-\fQ}{\hs_2}\form{\ba-\fQ}{\hs_3}\,.
\eeq

%----------------------------------------------------------------------------------------------
\section{Blocks computations}
\label{app:Blocks}
%----------------------------------------------------------------------------------------------

In this appendix, we summarize the essentials for the computations of the $\UU(1)$, $\bW_2$ and $\bW_3$ 3 and 4-point blocks as well as for the calculations of the blocks with insertions of the currents.

%----------------------------------------------------------------------------------------------
\subsection{The U(1) blocks}
\label{app:U1blocks}
%----------------------------------------------------------------------------------------------

We can define U(1) blocks in a fashion similar to the $\bW$ algebra case. The charge conservation seems built into the system. The current is $J_1(z)=i\partial \fb$, which has a mode expansion
\beq
J_1(z)=\sum_{n=-\infty}^{\infty}z^{-n-1}\am_n\,,\quad \text{ with }\quad \com{\am_n}{\am_m}=n\delta_{n+m,0}\,.
\eeq
The modes $\am_n$ form the $\hat{\mathfrak{u}}_1$ affine algebra.
We create representations by starting with $\ver_p$ annihilated by all $\am_n$ with $n>0$ that obeys $\am_0\ver_p=p\ver_p$. We are as generally in this article, denoting the vertex operator and the state it creates by the same symbol. Using the standard rule for the adjoint, we can define a Shapovalov form and find 
that  the norm of the state $\am_{-1}^{n_1}\ldots \am_{-m}^{n_m}\ver_p$ is given by $\prod_{j=1}^mn_j! j^{n_j}$. The numbers $n_j$ are related to the Young diagram $Y=\{Y_1,\ldots, Y_s\}$ as follows: the number $Y_j$ is the number of boxes of the $j^{\text{th}}$ row (drawn from the bottom upwards) of the Young diagram $Y$, while $n_r$ is number of rows in $Y$ of exactly $r$ boxes. For example, for $Y=\{1,1,2,4\}$ we have $n_1=2$, $n_2=1$, $n_3=0$ and $n_4=1$. 

We can compute as usual the recursion relations for the 3-point blocks
\beq
\label{eq:recursionrelationsforU1blocks}
\begin{split}
\vac{\ver_1(\infty)\ver_2(1)(\am_{-n}\widehat{\ver}_p)(0)}&=-\left(\delta_{n,0}p_1+p_2\right)\vac{\ver_1(\infty)\ver_2(1)\widehat{\ver}_p(0)}\,,\\ \bracket{\am_{-n}\widehat{\ver}_p}{\ver_3(1)\ver_4(0)}&=\left(p_3+\delta_{n,0}p_4\right)\bracket{\widehat{\ver}_p}{\ver_3(1)\ver_4(0)}\,,
\end{split}
\eeq
where $n\geq 0$. We remark that setting $n=0$ in the above, we obtain the charge conservation relations $p=-p_1-p_2$ for the first correlator and $p=p_3+p_4$ for the second. In general, we find that the 3-point blocks are given by $\gv_{12p}(\am_{-1}^{n_1}\ldots \am_{-m}^{n_m}\ver_p)=(-p_2)^{n_1+\cdots + n_m}$ and $\gvt_{p;34}(\am_{-1}^{n_1}\ldots \am_{-m}^{n_m}\ver_p)=p_3^{n_1+\cdots + n_m}$. It follows from the above discussion that the computation of the 4-point blocks factorizes leading to
\beq
\label{eq:U1blockgeneraldefinition}
\begin{split}
\block_{\UU(1)}&\equiv \block_{p}(p_1,p_2,p_3,p_4|q)=\sum_{n_1,n_2,\ldots =0}^{\infty}q^{\sum_{j=1}^\infty j n_j}\frac{(-p_2p_3)^{\sum_{r=1}^\infty n_r}}{\prod_{s=1}^\infty n_s! s^{n_s}}\\&=\prod_{j=1}^{\infty}\sum_{n=0}^{\infty}\frac{q^{j n}(-p_2p_3)^n}{n! j^n}=\prod_{j=1}^{\infty}e^{-\frac{p_2p_3q^j}{j}}=e^{\log(1-q)p_2p_3}=(1-q)^{p_2p_3}\,.
\end{split}
\eeq
We can now compute some conformal blocks with insertions of the current $J_1$.
We obtain after a short computation
 \beq
 \gvJ_{12p}(J_1(t);\am_{-1}^{n_1}\cdots \am_{-m}^{n_m}\ver_p)=\left\{\frac{p_2}{t-1}+\frac{p}{t}-\sum_{r=1}^m\frac{r n_r}{p_2}\frac{1}{t^{r+1}}\right\}(-p_2)^{n_1+\cdots +n_m}\,.
 \eeq
 After some computations, one finds from \eqref{eq:general4ptvvacFormula} the formula
 \beq
 \label{eq:vvacJ1for four points}
 \vvac{J_1(t)}_4=\frac{p_2}{t-1}+\frac{p}{t}-\frac{1}{p_2}\frac{(-p_2p_3)q}{t(t-q)}=\frac{p_2}{t-1}+\frac{p_3}{t-q}+\frac{p_4}{t}=\frac{\vac{J_1(t)\ver_{p_1}(\infty)\ver_{p_2}(1)\ver_{p_3}(q)\ver_{p_4}(0)}}{\vac{\ver_{p_1}(\infty)\ver_{p_2}(1)\ver_{p_3}(q)\ver_{p_4}(0)}}\,,
 \eeq
where we remind that $p=p_3+p_4$. 
We remark that $\vvac{J_1(t)}_4$ is equal to the ratio of the full correlation functions only for the $\UU(1)$ case because in that case we have charge conservation! This means that only one primary propagates in the four point function and therefore the structure constants cancel in the ratio.

%----------------------------------------------------------------------------------------------
\subsection{The Virasoro blocks}
\label{app: virasoro}
%----------------------------------------------------------------------------------------------

\paragraph{Three points. } The case of the 3-point W-blocks is almost trivial since the $\vvac{W_s}_3$ are completely fixed by the $\bW_N$ Ward identities and the shortening properties of the simple punctures. 
For the Liouville case, since (we ignore the anti-holomorphic pieces), 
\beq
\begin{split}
\vac{\ver_1(z_1)\ver_2(z_2)\ver_3(z_3)}&=z_{12}^{\cd_3-\cd_1-\cd_2}z_{13}^{\cd_2-\cd_1-\cd_3}z_{23}^{\cd_1-\cd_2-\cd_3}\,,
\\ \vac{T(t)\ver_1(z_1)\ver_2(z_2)\ver_3(z_3)}&=\sum_{i=1}^3\left(\frac{\cd_i}{(t-z_i)^2}+\frac{\partial_{z_i}}{t-z_i}\right)\vac{\ver_1(z_1)\ver_2(z_2)\ver_3(z_3)}\,,
\end{split}
\eeq
we find after setting $z_1\rightarrow \infty, z_2\rightarrow 1, z_3\rightarrow0$ 
\beq
\label{eq:gvJ123 for T}
\gvJ_{123}(T(t);\emptyset)=\vvac{T(t)}_3=\frac{\vac{T(t)\ver_1(z_1)\ver_2(z_2)\ver_3(z_3)}}{\vac{\ver_1(z_1)\ver_2(z_2)\ver_3(z_3)}}=\frac{\cd_1 t(t-1)+\cd_2 t+\cd_3(1-t)}{ t^2(t-1)^2}\,.
\eeq
In general, we have the recursion relations
\beq
\label{eq:iterativeequationsforgamma}
\begin{split}
\vac{\ver_1(\infty)\ver_2(1)(L_{-n}\widehat{\ver}_{\cd})(0)}&=\left(\cd+n\cd_2-(1-\delta_{n,0})\cd_1\right)\vac{\ver_1(\infty)\ver_2(1)\widehat{\ver}_{\cd}(0)}\,,\\
\bracket{L_{-n}\widehat{\ver}_{\cd}}{\ver_3(1)\ver_4(0)}&=\left(\cd+n\cd_3-(1-\delta_{n,0})\cd_4\right)\bracket{\widehat{\ver}_{\cd}}{\ver_3(1)\ver_4(0)}\,.
\end{split}
\eeq
We also occasionally need the relations
\beq
\label{eq:Lm1inserted1}
\begin{split}
\vac{\ver_1(\infty)(L_{-1}\ver_2)(1)\widehat{\ver}_{\cd}(0)}&=(\cd_1-\cd_2-\cd)\vac{\ver_1(\infty)\ver_2(1)\widehat{\ver}_{\cd}(0)}\\
\bracket{\widehat{\ver}_\cd}{(L_{-1}\ver_3)(1)\ver_4(0)}&=\left(\cd-\cd_3-\cd_4\right)\bracket{\widehat{\ver}}{\ver_3(1)\ver_4(0)}\,.
\end{split}
\eeq

\paragraph{Four points.}  Let us compute $\vvac{T(t)}_4$ up to quadratic order in $q$. 
In the formula \eqref{eq:definition gvJ}, we have $\bY=\{Y\}$. If $Y$ is the empty partition, we reproduce \eqref{eq:gvJ123 for T} by using \eqref{eq:iterativeequationsforgamma}
\beq
\label{eq:expressionforgamma12Tempty}
\begin{split}
\gvJ_{12\cd}(T(t);\emptyset)&=\frac{1}{\vac{\ver_1(\infty)\ver_2(1)\ver_\cd(0)}}\sum_{n=-\infty}^{0}t^{-n-2}\vac{\ver_1(\infty)\ver_2(1)(L_{n}\ver_\cd)(0)}\\
&=t^{-2}\cd+\sum_{n=1}^{\infty}t^{n-2}\left(\cd+n\cd_2-\cd_1\right)=\frac{\cd_1 (t-1) t+\cd_2 t-\cd  (t-1)}{(t-1)^2 t^2}\,,
\end{split}
\eeq
where we have made use of \eqref{eq:iterativeequationsforgamma}. 
Similarly, we compute
\beq
\begin{split}
&\gvJ_{12\cd}(T(t);\{1\})=\frac{1}{\vac{\ver_1(\infty)\ver_2(1)\ver_\cd(0)}}\sum_{n=-\infty}^{1}t^{-n-2}\vac{\ver_1(\infty)\ver_2(1)(L_{n}L_{-1}\ver_\cd)(0)}\\
&=t^{-3}2\cd+t^{-2}(1+\cd)\frac{\vac{\ver_1(\infty)\ver_2(1)(L_{-1}\ver_\cd)(0)}}{\vac{\ver_1(\infty)\ver_2(1)\ver_\cd(0)}}+\sum_{n=1}^{\infty}t^{n-2}\frac{\vac{\ver_1(\infty)\ver_2(1)(L_{-n}L_{-1}\ver_\cd)(0)}}{\vac{\ver_1(\infty)\ver_2(1)\ver_\cd(0)}}\\
&=\frac{2 \cd }{t^3}+\frac{(\cd +1) (\cd -\cd_1+\cd_2)}{t^2}+\frac{\cd_2 (\cd -\cd_1+\cd_2)}{(t-1)^2}+\frac{(\cd -\cd_1+\cd_2) (\cd -\cd_1+\cd_2+1)}{t}\\&-\frac{(\cd -\cd_1+\cd_2) (\cd -\cd_1+\cd_2+1)}{t-1}\,.
\end{split}
\eeq
In the above we have used the commutation relations $\com{L_n}{L_m}=(n-m)L_{n+m}+\frac{c}{12}n(n^2-1)\delta_{n+m,0}$. 
We compute in a similar fashion
\beq
\begin{split}
&\gvJ_{12\cd}(T(t);\{2\})
=\frac{c+8 \cd }{2 t^4}+\frac{3 (\cd -\cd_1+\cd_2)}{t^3}+\frac{(\cd +2) (\cd -\cd_1+2 \cd_2)}{t^2}+\frac{\cd_2 (\cd -\cd_1+2 \cd_2)}{(t-1)^2}\\&+\frac{(\cd -\cd_1+\cd_2+2) (\cd -\cd_1+2 \cd_2)}{t}-\frac{(\cd -\cd_1+\cd_2+2) (\cd -\cd_1+2 \cd_2)}{t-1}\,,
\end{split}
\eeq
as well as
\beq
\begin{split}
&\gvJ_{12\cd}(T(t);\{1,1\})
=\frac{6 \cd }{t^4}+\frac{2 (2 \cd +1) (\cd -\cd_1+\cd_2)}{t^3}+\frac{(\cd +2) (\cd -\cd_1+\cd_2) (\cd -\cd_1+\cd_2+1)}{t^2}\\&+\frac{\cd_2 (\cd -\cd_1+\cd_2) (\cd -\cd_1+\cd_2+1)}{(t-1)^2}+\frac{(\cd -\cd_1+\cd_2) (\cd -\cd_1+\cd_2+1) (\cd -\cd_1+\cd_2+2)}{t}\\&-\frac{(\cd -\cd_1+\cd_2) (\cd -\cd_1+\cd_2+1) (\cd -\cd_1+\cd_2+2)}{t-1}\,.
\end{split}
\eeq
Putting everything together, we get
\beq
\label{eq:vvacTforN2}
\begin{split}
\vvac{T(t)}_4=&\frac{1}{\block_{\cd}(\cd_1,\cd_2,\cd_3,\cd_4|q)}\Bigg[\gvJ_{12\cd}(T(t);\emptyset)+q \gvJ_{12\cd}(T(t);\{1\})(\shap_{\cd}^{(1)})^{-1}\gvt_{\alpha;34}(\{1\})\\&+q^2\big(\gvJ_{12\cd}(T(t);\{2\}),\gvJ_{12\cd}(T(t);\{2\})\big)(\shap_{\cd}^{(2)})^{-1}\left(\begin{array}{c}\gvt_{\alpha;34}(\{2\})\\\gvt_{\alpha;34}(\{1,1\})\end{array}\right)\Bigg]+\mathcal{O}(q^3)\,,
\end{split}
\eeq
where the Shapovalov form is to be found in \eqref{eq:shapLiouvillelevel23}. Comparison of \eqref{eq:vvacTforN2} with the curve coefficient $\tilde{\cuco}_2^{(4)}$ (see \eqref{eq:curvecoefficientsSCQCDgeneralNandk} and \eqref{eq:shiftinkappa}) for $N>2$ shows a perfect agreement if the parameter identifications of section~\ref{subsec:parameter identification} are taken into account. The block in the denominator is easily computed by taking the definition \eqref{eq:definition of W - blocks} and using \eqref{eq:iterativeequationsforgamma}. It reads
\beq
\label{eq:Virasoro 4 point block general}
\begin{split}
&\block_{\cd}(\cd_1,\cd_2,\cd_3,\cd_4|q)=1+\frac{q (\cd -\cd_1+\cd_2) (\cd +\cd_3-\cd_4)}{2 \cd }\\&+q^2 \Bigg[(\cd +\cd_3-\cd_4) (\cd +\cd_3-\cd_4+1) \Bigg(\frac{\left(\frac{c}{2}+4 \cd \right) (\cd -\cd_1+\cd_2) (\cd -\cd_1+\cd_2+1)}{4 c \cd ^2+2 c \cd +32 \cd ^3-20 \cd ^2}\\&-\frac{6 \cd  (\cd -\cd_1+2 \cd_2)}{4 c \cd ^2+2 c \cd +32 \cd ^3-20 \cd ^2}\Bigg)+(\cd +2 \cd_3-\cd_4) \Bigg(\frac{\left(4 \cd ^2+2 \left(2 \cd ^2+2 \cd \right)\right) (\cd -\cd_1+2 \cd_2)}{4 c \cd ^2+2 c \cd +32 \cd ^3-20 \cd ^2}\\&-\frac{6 \cd  (\cd -\cd_1+\cd_2) (\cd -\cd_1+\cd_2+1)}{4 c \cd ^2+2 c \cd +32 \cd ^3-20 \cd ^2}\Bigg)\Bigg]+\mathcal{O}(q^3)\,.
\end{split}
\eeq

%----------------------------------------------------------------------------------------------
\subsection{The $\bW_3$-blocks}
\label{app: W3 blocks}
%----------------------------------------------------------------------------------------------

\paragraph{Ward identities. }  In the $\bW_3$ case, we have to use the shortening condition for $\ver_2$ in order to use the Ward identities to compute the 3-point block with an insertion of $W_3(t)\equiv W(t)$. The Ward identity that we want to use is (see 2.4 of \cite{Fateev:2007ab})
\beq
\label{eq:W3insertioninV1V2V3}
\vac{W(t)\ver_1(z_1)\cdots \ver_n(z_n)}=\sum_{k=1}^n\left(\frac{\charge_k}{(t-z_k)^3}+\frac{W_{-1;k}}{(t-z_k)^2}+\frac{W_{-2;k}}{t-z_k}\right)\vac{\ver_1(z_1)\cdots \ver_n(z_n)}\,,
\eeq
where  $\charge_k\equiv \charge_3(\balpha_k)$ with the charge $\charge_3(\ba)$ defined in \eqref{eq:defchargew3}. The action of $W_{-1}$ and $W_{-2}$ cannot in general be expressed via simple differential operators. 
Taking \eqref{eq:W3insertioninV1V2V3}, multiplying with $z^m$, $m=0,\ldots, 4$, integrating in $z$ over a contour encircling all the insertion points and using the fact that $W(t)\propto \frac{1}{t^6}$ for $t\rightarrow \infty$ gives five global Ward identities (see for example \cite{Belavin:2016qaa} starting from eq. (2.18) there). Thus, for the 3-point function, we have 5 identities and 6 unknowns, namely the correlation functions 
$\vac{W_{-1}\ver_1\ver_2\ver_3}$ $\vac{\ver_1W_{-1}\ver_2\ver_3}$, $\vac{\ver_1\ver_2W_{-1}\ver_3}$
 and similarly another three with insertions of $W_{-2}$ instead. We can thus solve for all of them except for $\vac{\ver_1W_{-1}\ver_2\ver_3}$. We can then get rid of $\vac{\ver_1W_{-1}\ver_2\ver_3}$ by using the fact that the primary field $\ver_2$ is semi-degenerate and that it has the null-vector $\big(W_{-1}-\frac{3w(\balpha_2)}{2\cd(\balpha_2)}L_{-1}\big)\ver_2=0$, so that
\beq
\begin{split}
\vvac{\ver_1(z_1)(W_{-1}\ver_2)(z_2)\ver_3(z_3)}&=\frac{3\charge_2}{2\cd_2}\frac{\partial}{\partial z_2 }\log[\vac{\ver_1(z_1)\ver_2(z_2)\ver_3(z_3)}]\longrightarrow \frac{3\charge_2(\cd_1-\cd_2-\cd_3)}{2\cd_2}\,,
\end{split}
 \eeq
 after setting $z_1, z_2, z_3$ to $\infty, 1, 0$.
Therefore using the Ward identities, \eqref{eq:W3insertioninV1V2V3} and the null vector, we find
\beq
\label{eq:3pointconformalblockwithWforW3general}
\begin{split}
\vvac{W(t)}_3=&\frac{\charge_3}{t^3}+\frac{2 \cd_2 (\charge_1+\charge_3)+\charge_2 (3 \cd_1-\cd_2-3 \cd_3)}{2 \cd_2 t^2}+\frac{\cd_2 (\charge_1+\charge_3)+\charge_2 (3 \cd_1-2 \cd_2-3 \cd_3)}{\cd_2 t}\\&+\frac{\charge_2 (2 \cd_2+3 \cd_3-3 \cd_1)-\cd_2 (\charge_1+\charge_3)}{\cd_2(t-1)}-\frac{3 \charge_2 (\cd_2+\cd_3-\cd_1)}{2 \cd_2 (t-1)^2}+\frac{\charge_2}{(t-1)^3}\,.
\end{split}
\eeq

\paragraph{3-point blocks.} We can derive recursion relations like \eqref{eq:recursionrelationWspart2} for more general simple punctures with $W_{-1}\ver_2=u L_{-1}\ver_2$ for some parameter $u$. We find for $n>0$ the identity
\beq
\label{eq: W3 recursionrelationWs}
\begin{split}
\vac{\ver_1(\infty)\ver_2(1)(W_{-n}\widehat{\ver}_{\bcharge})(0)}=&\vac{\ver_1(\infty)\ver_2(1)(W_{0}\widehat{\ver}_{\bcharge})(0)}\\&+\left[\charge_1-\frac{n(n-3)}{2}\charge_{2}+nu(\cd_1-\cd_2-\cd_{\bcharge})\right]\vac{\ver_1(\infty)\ver_2(1)\widehat{\ver}_{\bcharge}(0)}\,.
\end{split}
\eeq
The last element that we need are the $\gvt$ vertices. They can be computed through the following general relation for $n>0$
\beq
\label{eq: W3 recursionrelationWs2}
\begin{split}
\bracket{W_{-n}\widehat{\ver}_{\bcharge}}{\ver_3(1)\ver_4(0)}=&\left(\frac{n(n+3)}{2}w_3-w_4\right)\bracket{\widehat{\ver}_{\bcharge}}{\ver_3(1)\ver_4(0)}+\bracket{W_0\widehat{\ver}_{\bcharge}}{\ver_3(1)\ver_4(0)}\\&+n\bracket{\widehat{\ver}_{\bcharge}}{(W_{-1}\ver_3)(1)\ver_4(0)}\,.
\end{split}
\eeq
If $\ver_3$ is a special puncture, we can use $W_{-1}\ver_3=\tfrac{3w_3}{2\cd_3}L_{-1}\ver_3$ and the relation \eqref{eq:Lm1inserted1}
to compute the $\gvt$ vertices iteratively.

The blocks with insertion of currents can be computed with the recursion relations \eqref{eq: W3 recursionrelationWs} and \eqref{eq: W3 recursionrelationWs2}. If $\widehat{\ver}_{\bcharge}=\ver_3$ is a primary field (a full puncture for the 3-point case) and if $u=\tfrac{3w_2}{2\cd_2}$ (i.e.~if $\ver_2$ is the standard simple puncture), we find by using \eqref{eq: W3 recursionrelationWs} for the 3-point $\bW_3$-block with an insertion of the current $W(z)$ the expression
\beq
\gvJ_{123}(W(t);\emptyset)=\sum_{n=0}^{\infty}t^{n-3}\frac{\vac{\ver_1(\infty)\ver_2(1)(W_{-n}\ver_{3})(0)}}{\vac{\ver_1(\infty)\ver_2(1)\ver_{3}(0)}}=\vvac{W(t)}_3
\eeq
where $\vvac{W(t)}_3$ was computed via the Ward identities in
\eqref{eq:3pointconformalblockwithWforW3general}.

\paragraph{Four points.}Let us compute the first few order of $\vvac{W(t)}_4\equiv \vvac{W_3(t)}_4$. The $\bW_3$ algebra is presented in appendix~\ref{app:W3 algebra}. Together with the recursion relations it is straightforward to use a computer algebra program to compute
\begin{align}
\label{eq:3pointblockswithWforW3 part 1}
&\gvJ_{12\bw}(W(t);\{\emptyset,\{1\}\})=\frac{\cd ^2}{t^4}+\frac{2 \cd  \cd_2 (\cd -\cd_1+\cd_2)+2 \cd_2 \charge^2+\charge (2 \cd_2 \charge_1-\charge_2 (3 \cd -3 \cd_1+\cd_2))}{2 \cd_2 t^3}\nonumber\\
&+\frac{1}{4 \cd_2^2 t^2}\Big[4 \cd_2^2 \left(\cd  (\cd -\cd_1+\cd_2)+(\charge+\charge_1)^2\right)-2 \cd_2 \charge_2 (\charge+\charge_1) (6 \cd -6 \cd_1+2 \cd_2+3)\nonumber\\
&+\charge_2^2 (3 \cd -3 \cd_1+\cd_2) (3 \cd -3 \cd_1+\cd_2+3)\Big]+\frac{1}{2 \cd_2^2 t}\Big[2 \cd_2^2 \left(\cd  (\cd -\cd_1+\cd_2)+(\charge+\charge_1)^2\right)\nonumber\\
&-\cd_2 \charge_2 (\charge+\charge_1) (9 \cd -9 \cd_1+5 \cd_2+6)+\charge_2^2 (3 \cd -3 \cd_1+\cd_2) (3 \cd -3 \cd_1+2 \cd_2+3)\Big]\\
&+\frac{1}{2 \cd_2^2 (t-1)}\Big[-2 \cd_2^2 \left(\cd  (\cd -\cd_1+\cd_2)+(\charge+\charge_1)^2\right)+\cd_2 \charge_2 (\charge+\charge_1) (9 \cd -9 \cd_1+5 \cd_2+6)\nonumber\\
&-\charge_2^2 (3 \cd -3 \cd_1+\cd_2) (3 \cd -3 \cd_1+2 \cd_2+3)\Big]\nonumber\\
&+\frac{3 \charge_2 (\cd -\cd_1+\cd_2+1) (\charge_2 (3 \cd -3 \cd_1+\cd_2)-2 \cd_2 (\charge+\charge_1))}{4 \cd_2^2 (t-1)^2}-\frac{\charge_2 (\charge_2 (3 \cd -3 \cd_1+\cd_2)-2 \cd_2 (\charge+\charge_1))}{2 \cd_2 (t-1)^3}\nonumber\,,
\end{align}
as well as
\beq
\label{eq:3pointblockswithWforW3 part 2}
\begin{split}
&\gvJ_{12\bw}(W(t);\{\{1\},\emptyset\}) =\frac{3 \cd}{t^4}+\frac{\cd_2 (\cd (\cd -\cd_1+\cd_2+2)+2 \charge_1)-\charge_2 (3 \cd -3 \cd_1+\cd_2)}{\cd_2 t^3}\\&+\frac{2 \cd_2 (\cd+\charge_1) (\cd -\cd_1+\cd_2+2)-\charge_2 \left(\cd_2 (4 \cd -4 \cd_1+5)+3 (\cd -\cd_1) (\cd -\cd_1+3)+\cd_2^2\right)}{2 \cd_2 t^2}\\&+\frac{(\cd -\cd_1+\cd_2+2) (\cd_2 (\cd+\charge_1)+\charge_2 (-3 \cd +3 \cd_1-2 \cd_2))}{\cd_2 t}+\frac{\charge_2 (\cd -\cd_1+\cd_2)}{(t-1)^3}\\&+\frac{(\cd -\cd_1+\cd_2+2) (\charge_2 (3 \cd -3 \cd_1+2 \cd_2)-\cd_2 (\charge+\charge_1))}{\cd_2 (t-1)}-\frac{3 \charge_2 (\cd -\cd_1+\cd_2) (\cd -\cd_1+\cd_2+1)}{2 \cd_2 (t-1)^2}\,.
\end{split}
\eeq
In \eqref{eq:3pointblockswithWforW3 part 1} and \eqref{eq:3pointblockswithWforW3 part 2}, we have put $Q=0$ from which follows $c=2$ and $\beta=\tfrac{1}{2}$ for simplicity. 

The four point block $\block$ to linear order in $q$ (for $Q\neq 0$) can be obtained quite straightforwardly by inverting \eqref{eq:W3shaplevel1} and using $\gv_{12\bw}(\{\{1\},\emptyset\})=\cd -\cd_1+\cd_2$, $\gv_{12\bw}(\{\emptyset,\{1\}\})=w+\charge_1+\frac{3 \charge_2 (-\cd +\cd_1-\cd_2)}{2 \cd_2}+\charge_2$, $\gvt_{\bw;34}(\{\{1\},\emptyset\})=\cd +\cd_3-\cd_4$, $\gvt_{\bw;34}(\{\emptyset,\{1\}\})=w+\frac{3 \charge_3 (\cd -\cd_3-\cd_4)}{2 \cd_3}+2 \charge_3-\charge_4$.  Thus, the 4-point $\bW_3$-block reads
\beq
\begin{split}
&\block_{\bw}(\bw_1,\bw_2,\bw_3,\bw_4|q)=1+\frac{q}{2 \cd_2 \left(27 \charge^2+\cd ^2 \left(4 \cd -3 Q^2\right)\right)}\\&\times \Bigg[\frac{3 (\cd_2 (\charge (\cd -3 \cd_1+3 \cd_2)-2 \cd  \charge_1)+\cd  \charge_2 (3 \cd -3 \cd_1+\cd_2)) (2 \cd_3 (\charge-\charge_4)+\charge_3 (3 \cd +\cd_3-3 \cd_4))}{\cd_3}\\&+(\cd +\cd_3-\cd_4) \left(18 \cd_2 \charge^2-9 \charge (\charge_2 (3 \cd -3 \cd_1+\cd_2)-2 \cd_2 \charge_1)-\cd  \cd_2 \left(3 Q^2-4 \cd \right) (\cd -\cd_1+\cd_2)\right)\Bigg]\\&+\mathcal{O}(q^2)\,.
\end{split}
\eeq
The higher orders in $q$ are computed similarly. Combining the block $\block$ with \eqref{eq:3pointblockswithWforW3 part 1} and \eqref{eq:3pointblockswithWforW3 part 2}, one can easily compute $\vvac{W(t)}_4$ to linear order in $q$.

%----------------------------------------------------------------------------------------------
\section{Instanton Partition Functions}
\label{app: Instanton Partition Functions}
%----------------------------------------------------------------------------------------------

For the $\SU(N)$ instanton partition functions, we define\footnote{See \cite{Tachikawa:2014dja} for a review. Our definition of the antifundamental partition function differs by a sign.}  $\epsilon=\epsilon_1+ \epsilon_2$ and consider first the matter contributions to the instanton partition function:
\beq
\begin{split}
\calZ_{\text{fund}}(\textbf{a},\bY;m)&=\prod_{s=1}^N\prod_{(i,j) \in Y_s}\Big[a_s+\epsilon_1i+\epsilon_2j-m \Big]\,,\\
\calZ_{\text{antifund}}(\textbf{a},\bY;m)&=\prod_{s=1}^N\prod_{(i,j) \in Y_s}\Big[\epsilon-m-a_s-\epsilon_1i-\epsilon_2j \Big]\,,\\
\calZ_{\text{bifund}}(\textbf{a},\bY;\textbf{a}',\bY';m)&=\prod_{s,s'=1}^N\prod_{(i,j)\in Y_s}\left[a_s-a'_{s'}-\epsilon_1L_{Y'_{s'}}(i,j)+\epsilon_2\left(A_{Y_s}(i,j)+1\right)-m\right]\\&\times \prod_{(i',j')\in Y'_{s'}}\left[\epsilon+a_s-a'_{s'}+\epsilon_1L_{Y_{s}}(i',j')-\epsilon_2\left(A_{Y'_{s'}}(i',j')+1\right)-m\right]\,,
\end{split}
\eeq
where we define the arm and leg lengths as
\beq
A_Y(i,j)=Y_i-j\,,\qquad L_{Y}(i,j)=Y^t_j-i\,.
\eeq
Finally, we have the vector multiplet contribution
\beq
\calZ_{\text{vec}}(\textbf{a},\bY)=\frac{1}{\calZ_{\text{bifund}}(\textbf{a},\bY;\textbf{a},\bY;0)}\,.
\eeq
Specializations of the bifundamental contribution lead to the following identities
\beq
\begin{split}
\calZ_{\text{bifund}}(\textbf{a},\bY;\textbf{b},\boldsymbol{\emptyset};m)&=\prod_{s=1}^N\calZ_{\text{fund}}(\textbf{a},\bY;m+b_s)\,,\\ 
\calZ_{\text{bifund}}(\textbf{b},\boldsymbol{\emptyset};\textbf{a},\bY;m)&=\prod_{s=1}^N\calZ_{\text{antifund}}(\textbf{a},\bY;m-b_s)\,.
\end{split}
\eeq

%----------------------------------------------------------------------------------------------
%Bibliography

\providecommand{\href}[2]{#2}\begingroup\raggedright\endgroup


\begin{thebibliography}{10}

\bibitem{Intriligator:1995au}
K.~A. Intriligator and N.~Seiberg, {\it {Lectures on supersymmetric gauge
  theories and electric-magnetic duality}},  {\em Nucl. Phys. Proc. Suppl.}
  {\bf 45BC} (1996) 1--28, [\href{http://arxiv.org/abs/hep-th/9509066}{{\tt
  hep-th/9509066}}]. [,157(1995)].

\bibitem{Seiberg:1994rs}
N.~Seiberg and E.~Witten, {\it {Electric - magnetic duality, monopole
  condensation, and confinement in N=2 supersymmetric Yang-Mills theory}},
  {\em Nucl. Phys.} {\bf B426} (1994) 19--52,
  [\href{http://arxiv.org/abs/hep-th/9407087}{{\tt hep-th/9407087}}]. [Erratum:
  Nucl. Phys.B430,485(1994)].

\bibitem{Seiberg:1994aj}
N.~Seiberg and E.~Witten, {\it {Monopoles, duality and chiral symmetry breaking
  in N=2 supersymmetric QCD}},  {\em Nucl. Phys.} {\bf B431} (1994) 484--550,
  [\href{http://arxiv.org/abs/hep-th/9408099}{{\tt hep-th/9408099}}].

\bibitem{Intriligator:1994sm}
K.~A. Intriligator and N.~Seiberg, {\it {Phases of N=1 supersymmetric gauge
  theories in four-dimensions}},  {\em Nucl. Phys.} {\bf B431} (1994) 551--568,
  [\href{http://arxiv.org/abs/hep-th/9408155}{{\tt hep-th/9408155}}].

\bibitem{Vafa:1997mh}
C.~Vafa, {\it {Geometric origin of Montonen-Olive duality}},  {\em Adv. Theor.
  Math. Phys.} {\bf 1} (1998) 158--166,
  [\href{http://arxiv.org/abs/hep-th/9707131}{{\tt hep-th/9707131}}].

\bibitem{Gaiotto:2009we}
D.~Gaiotto, {\it {N=2 dualities}},  {\em JHEP} {\bf 1208} (2012) 034,
  [\href{http://arxiv.org/abs/0904.2715}{{\tt arXiv:0904.2715}}].

\bibitem{Gaiotto:2009hg}
D.~Gaiotto, G.~W. Moore, and A.~Neitzke, {\it {Wall-crossing, Hitchin Systems,
  and the WKB Approximation}},  \href{http://arxiv.org/abs/0907.3987}{{\tt
  arXiv:0907.3987}}.

\bibitem{Hama:2012bg}
N.~Hama and K.~Hosomichi, {\it {Seiberg-Witten Theories on Ellipsoids}},  {\em
  JHEP} {\bf 09} (2012) 033, [\href{http://arxiv.org/abs/1206.6359}{{\tt
  arXiv:1206.6359}}]. [Addendum: JHEP10,051(2012)].

\bibitem{Pestun:2007rz}
V.~Pestun, {\it {Localization of gauge theory on a four-sphere and
  supersymmetric Wilson loops}},  {\em Commun.Math.Phys.} {\bf 313} (2012)
  71--129, [\href{http://arxiv.org/abs/0712.2824}{{\tt arXiv:0712.2824}}].

\bibitem{Alday:2009aq}
L.~F. Alday, D.~Gaiotto, and Y.~Tachikawa, {\it {Liouville Correlation
  Functions from Four-dimensional Gauge Theories}},  {\em Lett.Math.Phys.} {\bf
  91} (2010) 167--197, [\href{http://arxiv.org/abs/0906.3219}{{\tt
  arXiv:0906.3219}}].

\bibitem{Wyllard:2009hg}
N.~Wyllard, {\it {A(N-1) conformal Toda field theory correlation functions from
  conformal N = 2 SU(N) quiver gauge theories}},  {\em JHEP} {\bf 0911} (2009)
  002, [\href{http://arxiv.org/abs/0907.2189}{{\tt arXiv:0907.2189}}].

\bibitem{Teschner:2014oja}
J.~Teschner, {\it {Exact Results on ${\mathcal N}=2$ Supersymmetric Gauge
  Theories}},  in {\em New Dualities of Supersymmetric Gauge Theories}
  (J.~Teschner, ed.), pp.~1--30.
\newblock 2016.
\newblock \href{http://arxiv.org/abs/1412.7145}{{\tt arXiv:1412.7145}}.

\bibitem{Pestun:2016jze}
V.~Pestun and M.~Zabzine, {\it {Introduction to localization in quantum field
  theory}},  \href{http://arxiv.org/abs/1608.02953}{{\tt arXiv:1608.02953}}.

\bibitem{Leigh:1995ep}
R.~G. Leigh and M.~J. Strassler, {\it {Exactly marginal operators and duality
  in four-dimensional N=1 supersymmetric gauge theory}},  {\em Nucl. Phys.}
  {\bf B447} (1995) 95--136, [\href{http://arxiv.org/abs/hep-th/9503121}{{\tt
  hep-th/9503121}}].

\bibitem{Green:2010da}
D.~Green, Z.~Komargodski, N.~Seiberg, Y.~Tachikawa, and B.~Wecht, {\it {Exactly
  Marginal Deformations and Global Symmetries}},  {\em JHEP} {\bf 06} (2010)
  106, [\href{http://arxiv.org/abs/1005.3546}{{\tt arXiv:1005.3546}}].

\bibitem{Kachru:1998ys}
S.~Kachru and E.~Silverstein, {\it {4-D conformal theories and strings on
  orbifolds}},  {\em Phys. Rev. Lett.} {\bf 80} (1998) 4855--4858,
  [\href{http://arxiv.org/abs/hep-th/9802183}{{\tt hep-th/9802183}}].

\bibitem{Lawrence:1998ja}
A.~E. Lawrence, N.~Nekrasov, and C.~Vafa, {\it {On conformal field theories in
  four-dimensions}},  {\em Nucl. Phys.} {\bf B533} (1998) 199--209,
  [\href{http://arxiv.org/abs/hep-th/9803015}{{\tt hep-th/9803015}}].

\bibitem{Heckman:2016xdl}
J.~J. Heckman, P.~Jefferson, T.~Rudelius, and C.~Vafa, {\it {Punctures for
  Theories of Class $\mathcal{S}_\Gamma$}},
  \href{http://arxiv.org/abs/1609.01281}{{\tt arXiv:1609.01281}}.

\bibitem{Apruzzi:2016nfr}
F.~Apruzzi, F.~Hassler, J.~J. Heckman, and I.~V. Melnikov, {\it {From 6D SCFTs
  to Dynamic GLSMs}},  \href{http://arxiv.org/abs/1610.00718}{{\tt
  arXiv:1610.00718}}.

\bibitem{Gaiotto:2015usa}
D.~Gaiotto and S.~S. Razamat, {\it {$ \mathcal{N}=1 $ theories of class $
  {\mathcal{S}}_k $}},  {\em JHEP} {\bf 07} (2015) 073,
  [\href{http://arxiv.org/abs/1503.05159}{{\tt arXiv:1503.05159}}].

\bibitem{Franco:2015jna}
S.~Franco, H.~Hayashi, and A.~Uranga, {\it {Charting Class $\mathcal S_k$
  Territory}},  {\em Phys. Rev.} {\bf D92} (2015), no.~4 045004,
  [\href{http://arxiv.org/abs/1504.05988}{{\tt arXiv:1504.05988}}].

\bibitem{Hanany:2015pfa}
A.~Hanany and K.~Maruyoshi, {\it {Chiral theories of class $ \mathcal{S} $}},
  {\em JHEP} {\bf 12} (2015) 080, [\href{http://arxiv.org/abs/1505.05053}{{\tt
  arXiv:1505.05053}}].

\bibitem{Morrison:2016nrt}
D.~R. Morrison and C.~Vafa, {\it {F-theory and $ \mathcal{N} $ = 1 SCFTs in
  four dimensions}},  {\em JHEP} {\bf 08} (2016) 070,
  [\href{http://arxiv.org/abs/1604.03560}{{\tt arXiv:1604.03560}}].

\bibitem{Razamat:2016dpl}
S.~S. Razamat, C.~Vafa, and G.~Zafrir, {\it {4d N=1 from 6d (1,0)}},
  \href{http://arxiv.org/abs/1610.09178}{{\tt arXiv:1610.09178}}.

\bibitem{Pal:2016eqy}
S.~Pal and J.~Song, {\it {New Dualities and Misleading Anomaly Matchings from
  Outer-automorphism Twists}},  \href{http://arxiv.org/abs/1611.00694}{{\tt
  arXiv:1611.00694}}.

\bibitem{Garcia-Etxebarria:2016bpb}
I.~Garc\'ia-Etxebarria and B.~Heidenreich, {\it {S-duality in $\mathcal{N} = 1$
  orientifold SCFTs}},  \href{http://arxiv.org/abs/1612.00853}{{\tt
  arXiv:1612.00853}}.

\bibitem{Bah:2017gph}
I.~Bah, A.~Hanany, K.~Maruyoshi, S.~S. Razamat, Y.~Tachikawa, and G.~Zafrir,
  {\it {4d N=1 from 6d N=(1,0) on a torus with fluxes}},
  \href{http://arxiv.org/abs/1702.04740}{{\tt arXiv:1702.04740}}.

\bibitem{Coman:2015bqq}
I.~Coman, E.~Pomoni, M.~Taki, and F.~Yagi, {\it {Spectral curves of
  $\mathcal{N}=1$ theories of class $\mathcal{S}_k$}},
  \href{http://arxiv.org/abs/1512.06079}{{\tt arXiv:1512.06079}}.

\bibitem{Witten:1997sc}
E.~Witten, {\it {Solutions of four-dimensional field theories via M theory}},
  {\em Nucl.Phys.} {\bf B500} (1997) 3--42,
  [\href{http://arxiv.org/abs/hep-th/9703166}{{\tt hep-th/9703166}}].

\bibitem{Aganagic:2006ex}
M.~Aganagic, C.~Beem, J.~Seo, and C.~Vafa, {\it {Geometrically Induced
  Metastability and Holography}},  {\em Nucl. Phys.} {\bf B789} (2008)
  382--412, [\href{http://arxiv.org/abs/hep-th/0610249}{{\tt hep-th/0610249}}].

\bibitem{Aganagic:2008qa}
M.~Aganagic, C.~Beem, J.~Seo, and C.~Vafa, {\it {Extended Supersymmetric Moduli
  Space and a SUSY/Non-SUSY Duality}},  {\em Nucl. Phys.} {\bf B822} (2009)
  135--171, [\href{http://arxiv.org/abs/0804.2489}{{\tt arXiv:0804.2489}}].

\bibitem{Kozcaz:2010af}
C.~Kozcaz, S.~Pasquetti, and N.~Wyllard, {\it {A \& B model approaches to
  surface operators and Toda theories}},  {\em JHEP} {\bf 08} (2010) 042,
  [\href{http://arxiv.org/abs/1004.2025}{{\tt arXiv:1004.2025}}].

\bibitem{Keller:2011ek}
C.~A. Keller, N.~Mekareeya, J.~Song, and Y.~Tachikawa, {\it {The ABCDEFG of
  Instantons and W-algebras}},  {\em JHEP} {\bf 03} (2012) 045,
  [\href{http://arxiv.org/abs/1111.5624}{{\tt arXiv:1111.5624}}].

\bibitem{Lykken:1997gy}
J.~D. Lykken, E.~Poppitz, and S.~P. Trivedi, {\it {Chiral gauge theories from
  D-branes}},  {\em Phys. Lett.} {\bf B416} (1998) 286--294,
  [\href{http://arxiv.org/abs/hep-th/9708134}{{\tt hep-th/9708134}}].

\bibitem{Lykken:1997ub}
J.~D. Lykken, E.~Poppitz, and S.~P. Trivedi, {\it {M(ore) on chiral gauge
  theories from D-branes}},  {\em Nucl. Phys.} {\bf B520} (1998) 51--74,
  [\href{http://arxiv.org/abs/hep-th/9712193}{{\tt hep-th/9712193}}].

\bibitem{Bao:2011rc}
L.~Bao, E.~Pomoni, M.~Taki, and F.~Yagi, {\it {M5-Branes, Toric Diagrams and
  Gauge Theory Duality}},  {\em JHEP} {\bf 1204} (2012) 105,
  [\href{http://arxiv.org/abs/1112.5228}{{\tt arXiv:1112.5228}}].

\bibitem{Mitev:2014isa}
V.~Mitev and E.~Pomoni, {\it {Toda 3-Point Functions From Topological
  Strings}},  {\em JHEP} {\bf 06} (2015) 049,
  [\href{http://arxiv.org/abs/1409.6313}{{\tt arXiv:1409.6313}}].

\bibitem{Fateev:2007ab}
V.~Fateev and A.~Litvinov, {\it {Correlation functions in conformal Toda field
  theory. I.}},  {\em JHEP} {\bf 0711} (2007) 002,
  [\href{http://arxiv.org/abs/0709.3806}{{\tt arXiv:0709.3806}}].

\bibitem{Fateev:2011hq}
V.~Fateev and A.~Litvinov, {\it {Integrable structure, W-symmetry and AGT
  relation}},  {\em JHEP} {\bf 1201} (2012) 051,
  [\href{http://arxiv.org/abs/1109.4042}{{\tt arXiv:1109.4042}}].

\bibitem{Kanno:2009ga}
S.~Kanno, Y.~Matsuo, S.~Shiba, and Y.~Tachikawa, {\it {N=2 gauge theories and
  degenerate fields of Toda theory}},  {\em Phys. Rev.} {\bf D81} (2010)
  046004, [\href{http://arxiv.org/abs/0911.4787}{{\tt arXiv:0911.4787}}].

\bibitem{Fateev:1987zh}
V.~A. Fateev and S.~L. Lukyanov, {\it {The Models of Two-Dimensional Conformal
  Quantum Field Theory with Z(n) Symmetry}},  {\em Int. J. Mod. Phys.} {\bf A3}
  (1988) 507.

\bibitem{Bouwknegt:1992wg}
P.~Bouwknegt and K.~Schoutens, {\it {W symmetry in conformal field theory}},
  {\em Phys. Rept.} {\bf 223} (1993) 183--276,
  [\href{http://arxiv.org/abs/hep-th/9210010}{{\tt hep-th/9210010}}].

\bibitem{Mironov:2009dr}
A.~Mironov, S.~Mironov, A.~Morozov, and A.~Morozov, {\it {CFT exercises for the
  needs of AGT}},  \href{http://arxiv.org/abs/0908.2064}{{\tt
  arXiv:0908.2064}}.

\bibitem{Blumenhagen:1990jv}
R.~Blumenhagen, M.~Flohr, A.~Kliem, W.~Nahm, A.~Recknagel, and R.~Varnhagen,
  {\it {W algebras with two and three generators}},  {\em Nucl. Phys.} {\bf
  B361} (1991) 255--289.

\bibitem{Carstensen}
J.~P. Carstensen, V.~Mitev, and E.~Pomoni {\em {To appear}}.

\bibitem{Gerchkovitz:2014gta}
E.~Gerchkovitz, J.~Gomis, and Z.~Komargodski, {\it {Sphere Partition Functions
  and the Zamolodchikov Metric}},  {\em JHEP} {\bf 11} (2014) 001,
  [\href{http://arxiv.org/abs/1405.7271}{{\tt arXiv:1405.7271}}].

\bibitem{Bobev:2013cja}
N.~Bobev, H.~Elvang, D.~Z. Freedman, and S.~S. Pufu, {\it {Holography for $N =
  2^*$ on $S^4$}},  {\em JHEP} {\bf 07} (2014) 001,
  [\href{http://arxiv.org/abs/1311.1508}{{\tt arXiv:1311.1508}}].

\bibitem{Bobev:2016nua}
N.~Bobev, H.~Elvang, U.~Kol, T.~Olson, and S.~S. Pufu, {\it {Holography for
  $\mathcal{N}=1^*$ on $S^4$}},  \href{http://arxiv.org/abs/1605.00656}{{\tt
  arXiv:1605.00656}}.

\bibitem{Bershadsky:1998mb}
M.~Bershadsky, Z.~Kakushadze, and C.~Vafa, {\it {String expansion as large N
  expansion of gauge theories}},  {\em Nucl. Phys.} {\bf B523} (1998) 59--72,
  [\href{http://arxiv.org/abs/hep-th/9803076}{{\tt hep-th/9803076}}].

\bibitem{Bershadsky:1998cb}
M.~Bershadsky and A.~Johansen, {\it {Large N limit of orbifold field
  theories}},  {\em Nucl. Phys.} {\bf B536} (1998) 141--148,
  [\href{http://arxiv.org/abs/hep-th/9803249}{{\tt hep-th/9803249}}].

\bibitem{Gadde:2009dj}
A.~Gadde, E.~Pomoni, and L.~Rastelli, {\it {The Veneziano Limit of N = 2
  Superconformal QCD: Towards the String Dual of N = 2 SU(N(c)) SYM with N(f) =
  2 N(c)}},  \href{http://arxiv.org/abs/0912.4918}{{\tt arXiv:0912.4918}}.

\bibitem{Gadde:2010zi}
A.~Gadde, E.~Pomoni, and L.~Rastelli, {\it {Spin Chains in N=2 Superconformal
  Theories: From the $Z_2$ Quiver to Superconformal QCD}},  {\em JHEP} {\bf 06}
  (2012) 107, [\href{http://arxiv.org/abs/1006.0015}{{\tt arXiv:1006.0015}}].

\bibitem{Liendo:2011xb}
P.~Liendo, E.~Pomoni, and L.~Rastelli, {\it {The Complete One-Loop Dilation
  Operator of N=2 SuperConformal QCD}},  {\em JHEP} {\bf 07} (2012) 003,
  [\href{http://arxiv.org/abs/1105.3972}{{\tt arXiv:1105.3972}}].

\bibitem{Pomoni:2011jj}
E.~Pomoni and C.~Sieg, {\it {From N=4 gauge theory to N=2 conformal QCD:
  three-loop mixing of scalar composite operators}},
  \href{http://arxiv.org/abs/1105.3487}{{\tt arXiv:1105.3487}}.

\bibitem{Pomoni:2013poa}
E.~Pomoni, {\it {Integrability in N=2 superconformal gauge theories}},  {\em
  Nucl. Phys.} {\bf B893} (2015) 21--53,
  [\href{http://arxiv.org/abs/1310.5709}{{\tt arXiv:1310.5709}}].

\bibitem{Mitev:2014yba}
V.~Mitev and E.~Pomoni, {\it {Exact effective couplings of four dimensional
  gauge theories with $\mathcal N=$ 2 supersymmetry}},  {\em Phys. Rev.} {\bf
  D92} (2015), no.~12 125034, [\href{http://arxiv.org/abs/1406.3629}{{\tt
  arXiv:1406.3629}}].

\bibitem{Mitev:2015oty}
V.~Mitev and E.~Pomoni, {\it {Exact Bremsstrahlung and Effective Couplings}},
  {\em JHEP} {\bf 06} (2016) 078, [\href{http://arxiv.org/abs/1511.02217}{{\tt
  arXiv:1511.02217}}].

\bibitem{Tom}
T.~Bourton and E.~Pomoni {\em {To appear}}.

\bibitem{Cordova:2016cmu}
C.~Cordova and D.~L. Jafferis, {\it {Toda Theory From Six Dimensions}},
  \href{http://arxiv.org/abs/1605.03997}{{\tt arXiv:1605.03997}}.

\bibitem{Dorey:2002ik}
N.~Dorey, T.~J. Hollowood, V.~V. Khoze, and M.~P. Mattis, {\it {The Calculus of
  many instantons}},  {\em Phys. Rept.} {\bf 371} (2002) 231--459,
  [\href{http://arxiv.org/abs/hep-th/0206063}{{\tt hep-th/0206063}}].

\bibitem{Nekrasov:2002qd}
N.~A. Nekrasov, {\it {Seiberg-Witten prepotential from instanton counting}},
  {\em Adv. Theor. Math. Phys.} {\bf 7} (2003), no.~5 831--864,
  [\href{http://arxiv.org/abs/hep-th/0206161}{{\tt hep-th/0206161}}].

\bibitem{Nekrasov:2003rj}
N.~Nekrasov and A.~Okounkov, {\it {Seiberg-Witten theory and random
  partitions}},  {\em Prog. Math.} {\bf 244} (2006) 525--596,
  [\href{http://arxiv.org/abs/hep-th/0306238}{{\tt hep-th/0306238}}].

\bibitem{Nekrasov:2016ydq}
N.~Nekrasov, {\it {BPS/CFT Correspondence III: Gauge Origami partition function
  and qq-characters}},  \href{http://arxiv.org/abs/1701.00189}{{\tt
  arXiv:1701.00189}}.

\bibitem{Belavin:2011pp}
V.~Belavin and B.~Feigin, {\it {Super Liouville conformal blocks from N=2 SU(2)
  quiver gauge theories}},  {\em JHEP} {\bf 07} (2011) 079,
  [\href{http://arxiv.org/abs/1105.5800}{{\tt arXiv:1105.5800}}].

\bibitem{Nishioka:2011jk}
T.~Nishioka and Y.~Tachikawa, {\it {Central charges of para-Liouville and Toda
  theories from M-5-branes}},  {\em Phys. Rev.} {\bf D84} (2011) 046009,
  [\href{http://arxiv.org/abs/1106.1172}{{\tt arXiv:1106.1172}}].

\bibitem{Bonelli:2011jx}
G.~Bonelli, K.~Maruyoshi, and A.~Tanzini, {\it {Instantons on ALE spaces and
  Super Liouville Conformal Field Theories}},  {\em JHEP} {\bf 08} (2011) 056,
  [\href{http://arxiv.org/abs/1106.2505}{{\tt arXiv:1106.2505}}].

\bibitem{Bonelli:2011kv}
G.~Bonelli, K.~Maruyoshi, and A.~Tanzini, {\it {Gauge Theories on ALE Space and
  Super Liouville Correlation Functions}},  {\em Lett. Math. Phys.} {\bf 101}
  (2012) 103--124, [\href{http://arxiv.org/abs/1107.4609}{{\tt
  arXiv:1107.4609}}].

\bibitem{Wyllard:2011mn}
N.~Wyllard, {\it {Coset conformal blocks and N=2 gauge theories}},
  \href{http://arxiv.org/abs/1109.4264}{{\tt arXiv:1109.4264}}.

\bibitem{Alfimov:2011ju}
M.~N. Alfimov and G.~M. Tarnopolsky, {\it {Parafermionic Liouville field theory
  and instantons on ALE spaces}},  {\em JHEP} {\bf 02} (2012) 036,
  [\href{http://arxiv.org/abs/1110.5628}{{\tt arXiv:1110.5628}}].

\bibitem{Belavin:2011sw}
A.~A. Belavin, M.~A. Bershtein, B.~L. Feigin, A.~V. Litvinov, and G.~M.
  Tarnopolsky, {\it {Instanton moduli spaces and bases in coset conformal field
  theory}},  {\em Commun. Math. Phys.} {\bf 319} (2013) 269--301,
  [\href{http://arxiv.org/abs/1111.2803}{{\tt arXiv:1111.2803}}].

\bibitem{Bao:2013pwa}
L.~Bao, V.~Mitev, E.~Pomoni, M.~Taki, and F.~Yagi, {\it {Non-Lagrangian
  Theories from Brane Junctions}},  {\em JHEP} {\bf 1401} (2014) 175,
  [\href{http://arxiv.org/abs/1310.3841}{{\tt arXiv:1310.3841}}].

\bibitem{Isachenkov:2014eya}
M.~Isachenkov, V.~Mitev, and E.~Pomoni, {\it {Toda 3-Point Functions From
  Topological Strings II}},  {\em JHEP} {\bf 08} (2016) 066,
  [\href{http://arxiv.org/abs/1412.3395}{{\tt arXiv:1412.3395}}].

\bibitem{Alday:2009fs}
L.~F. Alday, D.~Gaiotto, S.~Gukov, Y.~Tachikawa, and H.~Verlinde, {\it {Loop
  and surface operators in N=2 gauge theory and Liouville modular geometry}},
  {\em JHEP} {\bf 01} (2010) 113, [\href{http://arxiv.org/abs/0909.0945}{{\tt
  arXiv:0909.0945}}].

\bibitem{Okuda:2014fja}
T.~Okuda, {\it {Line operators in supersymmetric gauge theories and the 2d-4d
  relation}},  in {\em New Dualities of Supersymmetric Gauge Theories}
  (J.~Teschner, ed.), pp.~195--222.
\newblock 2016.
\newblock \href{http://arxiv.org/abs/1412.7126}{{\tt arXiv:1412.7126}}.

\bibitem{Gukov:2014gja}
S.~Gukov, {\it {Surface Operators}},  in {\em New Dualities of Supersymmetric
  Gauge Theories} (J.~Teschner, ed.), pp.~223--259.
\newblock 2016.
\newblock \href{http://arxiv.org/abs/1412.7127}{{\tt arXiv:1412.7127}}.

\bibitem{Ito:2016fpl}
Y.~Ito and Y.~Yoshida, {\it {Superconformal index with surface defects for
  class ${\cal S}_k$}},  \href{http://arxiv.org/abs/1606.01653}{{\tt
  arXiv:1606.01653}}.

\bibitem{Maruyoshi:2016caf}
K.~Maruyoshi and J.~Yagi, {\it {Surface defects as transfer matrices}},  {\em
  PTEP} {\bf 2016} (2016), no.~11 113B01,
  [\href{http://arxiv.org/abs/1606.01041}{{\tt arXiv:1606.01041}}].

\bibitem{Yagi:2017hmj}
J.~Yagi, {\it {Surface defects and elliptic quantum groups}},
  \href{http://arxiv.org/abs/1701.05562}{{\tt arXiv:1701.05562}}.

\bibitem{Costis}
J.~P. Carstensen, J.~A. Hayling, R.~Panerai, C.~Papageorgakis, and E.~Pomoni
  {\em {Work in Progress}}.

\bibitem{Belavin:2016qaa}
V.~Belavin, B.~Estienne, O.~Foda, and R.~Santachiara, {\it {Correlation
  functions with fusion-channel multiplicity in $ {\mathcal{W}}_3 $ Toda field
  theory}},  {\em JHEP} {\bf 06} (2016) 137,
  [\href{http://arxiv.org/abs/1602.03870}{{\tt arXiv:1602.03870}}].

\bibitem{Tachikawa:2014dja}
Y.~Tachikawa, {\it {A review on instanton counting and W-algebras}},  in {\em
  New Dualities of Supersymmetric Gauge Theories} (J.~Teschner, ed.),
  pp.~79--120.
\newblock 2016.
\newblock \href{http://arxiv.org/abs/1412.7121}{{\tt arXiv:1412.7121}}.

\end{thebibliography}
\end{document}